\documentclass{aa}  
\usepackage{graphicx}
\usepackage{txfonts}
\usepackage{natbib}
\bibpunct{(}{)}{;}{a}{}{,} % to follow the A&A style

\usepackage[bookmarksopen]{hyperref}

\newcommand{\xmm}{{\sl XMM-Newton}}
\newcommand{\cxo}{{\sl Chandra}}
\newcommand{\at}{AA\,Tau}
\newcommand{\lx}{L_{\rm X}}

\newcommand{\lxlbol}{L_{\rm X}/L_{\rm bol}}
\newcommand{\cnts}{\,counts\,s$^{-1}$}

\newcommand{\ergscm}{\,erg\,s$^{-1}$\,cm$^{-2}$}
\newcommand{\ergs}{\,erg\,s$^{-1}$}

\begin{document}
   \title{Observation of enhanced X-ray emission from the CTTS \at\\
during a transit of an accretion funnel
flow\thanks{Figures~\ref{fig:coup_lc}, \ref{fig:spectra} and
  \ref{fig:uv_x}, and Appendix~\ref{appendix:newabun} are only
  available in electronic form via {\tt
    http://www.edpsciences.org}\,.}}

\titlerunning{Enhanced X-ray emission from \at}

%   \subtitle{}

   \author{N.\ Grosso\inst{1}
\and J.\ Bouvier\inst{2}
\and T.\ Montmerle\inst{2}
\and M.\ Fern{\'a}ndez\inst{3}
\and K.\ Grankin\inst{4}
\and M.R.\ Zapatero Osorio\inst{5}
}
        
%\offprints{N.\ Grosso}

   \institute{Observatoire astronomique de Strasbourg, Universit{\'e}
              Louis-Pasteur, CNRS, INSU, 11 rue de l'Universit{\'e}, 67000
              Strasbourg, France
         \and Laboratoire d'astrophysique de Grenoble,
              Universit{\'e} Joseph-Fourier, CNRS, INSU,
	      414 rue de la Piscine,
              38041 Grenoble, France
         \and Instituto de Astrof{\'\i}sica de Andaluc{\'\i}a, CSIC, Camino Bajo
              de Hu{\'e}tor 50, 18008 Granada, Spain 
	 \and Ulugh Beg Astronomical Institute of the Uzbek Academy of
	 Sciences, Astronomicheskaya 33, 700052 Tashkent, Uzbekistan
         \and Instituto de Astrof{\'\i}sica de Canarias (IAC), v{\'\i}a L{\'a}ctea
              s/n, 38205 La Laguna, Tenerife, Spain}

   \date{Received 15 June 2007 / Accepted 27 August 2007}

% \abstract{}{}{}{}{} 
% 5 {} token are mandatory7

  \abstract
   {Classical T~Tauri stars are young solar-type stars accreting
  material from their circumstellar disks. Thanks to a favorable
  inclination of the system, the classical T Tauri star \at~exhibits
  periodic optical eclipses as the warped inner disk edge occults the
  stellar photosphere.}
   {We intend to observe the X-ray and UV
  emission of \at~during the optical eclipses with the aim to localize
  these  emitting regions on the star.}
   {\at~was observed for about 5\,h per \xmm~orbit (2 days) over 8
  successive orbits, which covers two optical eclipse periods
  (8.22 days). The \xmm~optical/UV monitor simultaneously provided UV
  photometry (UVW2 filter at 206\,nm) with a $\sim$15\,min sampling
  rate. Some $V$-band photometry was also obtained from the ground
  during this period in order to determine the dates of the eclipses.}
   {Two X-ray and UV measurements were secured close to the
  center of the eclipse ($\Delta V\sim1.5$\,mag). The UV
  flux is the highest just before the eclipse starts and the lowest
  towards the end of it. UV flux variations amount to a few 0.1\,mag
  on a few hours timescale, and up to 1\,mag on a week timescale, none
  of which are correlated with the X-ray flux. We model it with a
  weekly modulation (inner disk eclipse), plus a daily modulation,
  which suggests a non-steady accretion, but needs a longer
  observation to be confirmed. No such eclipses are detected in
  X-rays. Within each
  5\,h-long observations, \at~has a nearly constant X-ray count
  rate. On a timescale of days to weeks, the X-ray flux varies by a
  factor of 2--8, except for one measurement where the X-ray count rate
  was nearly 50 times stronger than the minimum observed level even
  though photoelectric absorption was the highest at this phase, and
  the plasma temperature reached 60\,MK, i.e. a factor of 2--3 higher
  than in the other observations. This X-ray event, observed close to
  the center of the optical eclipse, is interpreted as an X-ray
  flare. 
  }
   {We identify the variable column density with the low-density
  accretion funnel flows blanketing the magnetosphere. The lack of
  X-ray eclipses indicates that X-ray emitting regions are located at
  high latitudes. Furthermore, the occurrence of a strong X-ray flare
  near the center of the optical eclipse suggests that the
  magnetically active areas are closely associated with the base of
  the high-density accretion funnel flow. We speculate that the
  impact of this free falling accretion flow onto the strong magnetic
  field of the stellar corona may boost the X-ray emission.}

   \maketitle
%
%________________________________________________________________
%________________________________________________________________

\section{Introduction}

T~Tauri stars (TTSs), i.e.\ young (1--10\,Myrs) solar-type stars, are
conspicuous X-ray emitters. Their high X-ray luminosities ($\lx
\simeq 10^{28-31}$\,erg\,s$^{-1}$) compared to the Sun ($\lx\simeq
10^{27}$\,erg\,s$^{-1}$ at solar maximum), and intense flaring
activity (up to $\lx \simeq 10^{32-33}$\,erg\,s$^{-1}$) make them appear
as extremelly active young Suns in the X-ray domain. The analogy with
the solar activity has been quite successful in ascribing the X-ray
emission of TTSs to an optically thin, magnetically confined coronal plasma
in collisional equilibrium at temperatures of 10--100\,MK, emitting a
thermal bremsstrahlung continuum and emission lines \citep[see,
  e.g., review by ][]{feigelson99}. As observed in the X-ray coronae
of active stars \citep[e.g.,][]{ness04}, the enhanced X-ray luminosity
of TTSs can easily be explained by coronal structures with high plasma
density, because the X-ray luminosity is proportional to the plasma
emission measure, which scales linearly with the plasma volume but
with the square of the plasma electronic density. That most of TTSs
X-ray emission arises in an active magnetic corona is supported by
direct Zeeman measurements on photospheric spectral lines which
indicate surface magnetic fields of a few kilogauss
\citep[e.g.,][]{guenther99,johnskrull99,johnskrull04,johnskrull07,yang05}.
However, the dynamo mechanism producing the magnetic field in these
fully convective stars is still discussed
\citep[e.g.,][]{preibisch05b,briggs07}.

The solar paradigm, where the X-ray emitting plasma is confined in
magnetic loop with both feet anchored on the stellar photosphere, has been
questionned in the context of a Classical TTS (CTTS), which accretes
material from its circumstellar disk. Inside a few stellar radii above the
CTTS surface, the stellar magnetic field pressure is larger than
the ram pressure of the accreting gas. As a result, the stellar
magnetosphere truncates the inner accretion disk and controls the
accretion flows. The gas is mainly accreted from the disk edge to the
stellar surface along the dominant large scale stellar magnetic lines,
creating accretion funnel flows. The free-falling gas hits the stellar
surface at the feet of the accretion funnel flows, where the kinetic
energy is dissipated in a shock producing hot excess emission
\citep[see review on magnetospheric accretion by ][]{bouvier07b}.

The X-ray grating spectrometers aboard \cxo~and \xmm~are able to
obtain spectra of the X-ray brightest CTTSs, where the emission line
triplets of He-like elements are resolved, which provides a powerful tool
to assess the electronic density of the X-ray emitting plasma
\citep[e.g.,][]{porquet01}. In several CTTSs, plasma with high electronic
density ($n_{\rm e}\sim 10^{13}$\,cm$^{-2}$) and low temperature
($\sim$3\,MK), untypical of stellar coronae, were identified, and
therefore attributed to accretion shocks
\citep{kastner02,stelzer04c,schmitt05,robrade06,argiroffi07,guenther07,telleschi07};
whereas some CTTSs display no such evidence
\citep{audard05,smith05b,guedel07b}.

During the {\sl
 Chandra Orion Ultradeep Project} \citep[COUP, see][]{getman05b}, where
the TTSs of the Orion nebula cluster were monitored nearly
continuously over $\sim$13 days with the {\sl Advanced CCD Imaging
  Spectrometer}, numerous X-ray flares were observed. In a few cases,
 a size of several stellar radii was derived for the magnetic loop
 confining the X-ray emitting plasma, large enough to connect the stellar
surface with the edge of the accretion disk \citep{favata05}.

We propose to directly constrain the source and location of the X-ray
emission in CTTSs by using eclipses. Eclipse mapping have been
successfully used in binary active stars to reconstruct the coronae 
\citep[e.g.,][]{guedel01,guedel03b}, or to localize the flaring plasma
\citep[e.g.,][]{schmitt99,schmitt03}. 
Our target star is \at, located in the Taurus molecular cloud
  complex at a distance of $\sim$140\,pc \citep[e.g.,][]{kenyon94}. 
\at~is a quite typical member of the CTTS class, with a K7
  spectral type, a bolometric luminosity of $\sim$0.8\,L$_\odot$, a
  stellar mass of $\sim$0.8\,M$_\odot$, and a stellar radius of
  $\sim$1.85\,R$_\odot$; exhibiting moderate accretion disk
diagnostics (near-IR excess, optical veiling, Balmer line emission), 
but with the remarkable
property to be viewed nearly
edge-on. \citet{bouvier99,bouvier03,bouvier07} reported evidence for a
modulation of the photospheric flux and spectroscopic diagnostics with
a period of 8.22 days, corresponding to the rotational period of the
star. This stellar flux modulation was interpreted as the periodic
eclipse of the stellar photosphere by the optically thick,
magnetically-warp inner disk edge located at 8.8 stellar
radii. Photopolarimetric variations confirm the presence of an
optically thick wall located at the disk edge, and eclipsing
periodically the stellar photosphere \citep{menard03}.
 
We obtained 8 observations of \at~with \xmm~\citep{jansen01}, which
allows simultaneous observations with an EPIC pn \citep{strueder01}
and two EPIC MOS \citep{turner01} X-ray spectroimaging cameras, and
the Optical/UV monitor \citep[OM;][]{mason01}. We supplemented these
X-rays and UV observations with an optical ground-based monitoring of
\at~to secure the dates of the optical eclipses.
These \xmm~observations were previously reported in \citet[][hereafter
  SR]{schmitt07}, who used the minimum of the UV light curve as proxy
of the optical eclipse. SR found variable X-ray absorption ``such
that the times of maximal X-ray absorption and UV extinction coincide''.
SR introduced an additional absorption in a disk wind, or a peculiar
dust grain distribution to reconcile the high value of the X-ray
absorption outside the eclipse and the low optical extinction. 

In Sect.~\ref{uv_x_properties}, we present the X-ray and UV properties
of \at~based on a reanalysing of the full data set provided by these
\xmm~observations. In particular, we report X-ray (pn+MOS1+MOS2) and
UV light curves with a time resolution of $\sim$15\,min. We show that
a bright and hot flare was observed during the second observation.
In Sect.~\ref{variability}, thanks to our ground-based observations
and optical OM data, we determine the dates of the optical eclipses,
which allow us to compare the UV and X-ray variabilities versus the
rotational phase. We show that the UV minimum is outside the eclipse,
and that the (confirmed) variation of the column density are not
correlated with the rotational phase. In Sect.~\ref{discussion}, we
propose another origin for this variable column density, and we
discuss the origin of the X-ray flare that was observed during an
optical eclipse.

\begin{table}[!t]
\caption{Journal of the {\sl XMM-Newton} observations of \at~(PI: J.\ Bouvier).}
\label{table:journal}
\begin{tabular}{@{}cccccc@{}rc@{}rc@{}r@{}}
\hline
\hline
\noalign{\smallskip}
Obs. & Rev. & ObsId & Feb.\ 2003$^{\mathrm{a}}$ & Exposure$^{\mathrm{a}}$ \\
   & & & (d) & (h)  \\
\hline
\noalign{\smallskip}
1 & 583 & 015680201 & 14.10--14.29 & 4.7 \\
2 & 584 & 015680301 & 16.13--16.33 & 4.7 \\
3 & 585 & 015680501 & 18.07--18.26 & 4.7 \\
4 & 586 & 015680401 & 20.02--20.22 & 4.7 \\
5 & 587 & 015680601 & 22.13--22.35 & 5.4 \\
6 & 588 & 015680701 & 24.03--24.25 & 5.3 \\
7 & 589 & 015680801 & 26.52--26.71 & 4.6 \\
8 & 590 & 015680901 & 28.30--28.50 & 4.7 \\
\hline
\end{tabular}
\begin{list}{}{}
\item[$^{\mathrm{a}}$] The observation beginning, end, and duration is given for MOS1.
\end{list}
\end{table}

%________________________________________________________________
%________________________________________________________________
\section{X-ray and UV properties of \at}
\label{uv_x_properties}

\subsection{XMM-Newton observations and event selections}
\label{xmm}

Our observational strategy with \xmm~was to cover two consecutive
modulation periods ($\sim$17 days) in order to demonstrate the
reproducibility of the phenomenon from one rotational cycle to the
next. The temporal sampling of the X-ray light curve needs not to be
very dense because \at~spends nearly the same amount of time being
eclipsed as being entirely visible. Therefore, we requested one
4\,h-exposure with EPIC pn per \xmm~orbit (2 days) over 8 successive
orbits. We used the full frame science mode of the EPIC cameras with
the medium optical blocking filter. The pointing nominal coordinates were
$04^{\rm h}34^{\rm m}55\fs5$, $24\degr28\arcmin54\farcs0$ (J2000
equinox). The journal of the \xmm~observations of \at~is given in
Table~\ref{table:journal}.

The data reduction was made using the \xmm~Science Analysing
System (SAS, version 7.0.0)
For each observations, the event lists for each camera of EPIC were
produced using the SAS tasks {\tt epchain} and {\tt emchain},
respectively.  We used the background lightcurves
computed by these task in the 7.0--15.0\,keV energy range to determine
the time intervals affected by background proton flares. More than
half of the observing time is affected by
bad space weather. For each instrument, we made from the low
background time intervals a sky image with $3\arcsec$-pixels in the
0.5--7.3\,keV energy range\footnote{We selected single, double,
  triple, and quadruple pixel events (i.e.\ {\tt PATTERN} in the 0 to
  12 range), and also applied the predefined filter {\tt \#XMMEA\_EM}
  for MOS.}. The X-ray counterpart of \at~was detected in all the
observations.

\begin{figure*}[!t]
\centering
\includegraphics[angle=90,width=2\columnwidth]{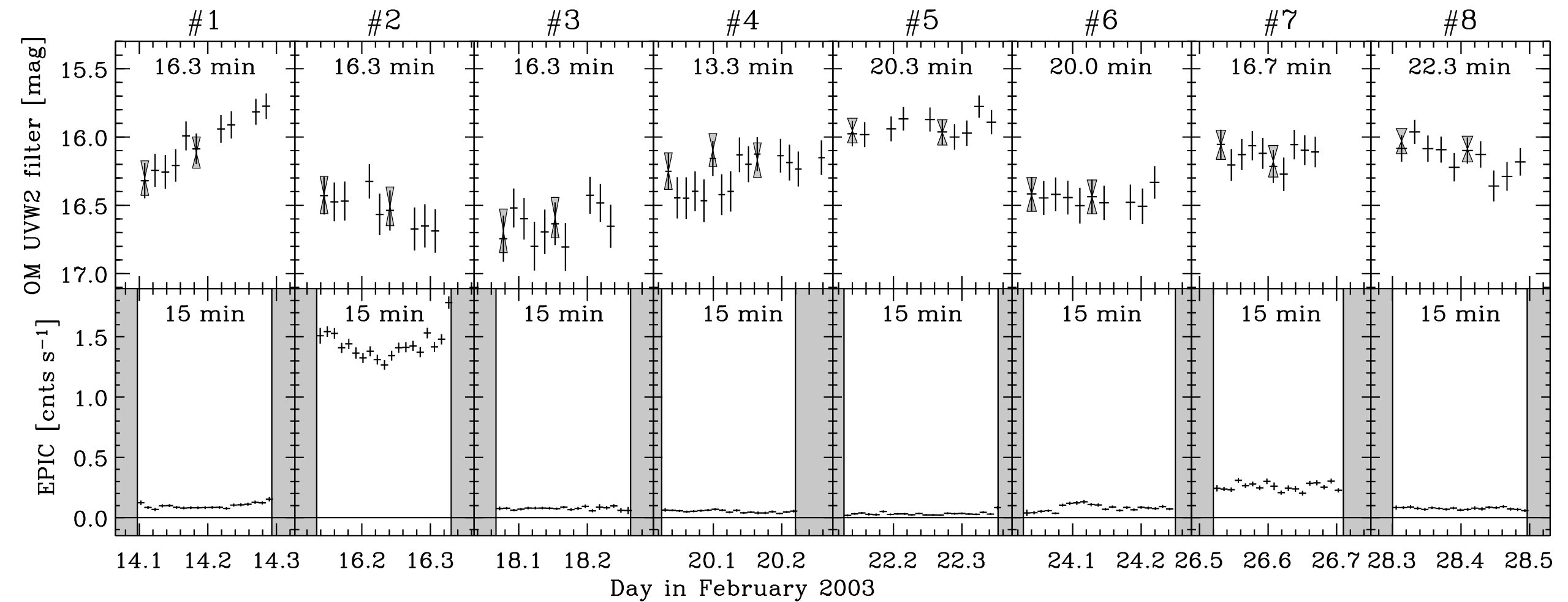}
 \caption{{\sl XMM-Newton} observations of \at. Top and bottom
   panels show the background subtracted UV (180--250\,nm) and X-ray
   (0.5--7.3\,keV) light
   curves obtained with the Optical/UV Monitor (OM) and EPIC
   (pn+MOS1+MOS2), respectively. In each panel, the label indicates
   the time interval used to bin the light curve. Grey hourglasses and
   crosses data points
   indicate UV photometry obtained with the large and small OM central
   imaging window. The X-ray count rates are corrected from (circular)
   aperture. The gray stripes indicate the beginning and the end of
   the EPIC observation.}
\label{fig:om_epic_lightcurves}
\end{figure*}

For the X-ray light curves of \at,
we selected  the source+background events within a circular region
centered on \at~. The extraction position and radius were optimized to
maximize the signal-to-noise ratio in the sky image. The background in
MOS and pn was extracted using an annular region centered on \at~and a
box region at the same distance to the CCD readout node, respectively,
where areas illuminated by other weak X-ray sources were excluded. 
For the X-ray spectra, we used the same extraction regions and only
time intervals with low background; we selected events with energy
above 0.3\,keV and the usual stronger selection criteria\footnote{For
  pn, we selected only single and double pixel events (i.e.\ {\tt
    PATTERN} in the 0 to 4 range) with {\tt FLAG} value equal to zero;
  for MOS, we selected {\tt PATTERN} in the 0 to 12 range, and applied
  the predefined filter {\tt \#XMMEA\_SM}.}. We computed the
corresponding redistribution matrix files and ancillary response files.

%________________________________________________________________
\subsection{X-ray light curves}
\label{lc}

For each instrument and observations, we first built the
source+background and the background light curves with 1\,s time bins
starting at the first good time interval (GTI) of MOS1. We rebinned
the light curve to 900\,s  to increase the signal. Then, we subtracted from the
source+background light curve the background light curve scaled to the
same source extraction area.

From the GTI extension, we scaled up with an IDL routine count rates
and errors affected by any lost of observing time, mainly due to the
triggering of counting mode during high flaring background periods,
when the count rate exceeded the detector telemetry limit. We also
corrected count rates and errors for circular aperture photometry using the
fraction of PSF counts inside the (circular) extraction region 
calculated by the SAS assuming a fixed photon energy of 1.5\,keV.
Finally, the light curves of the three detectors
were summed to produce the EPIC light curves. We estimated the missing
pn data at the beginning of the observations by multiplying the
MOS1+MOS2 count rates by 1.17, the median scaling factor between
MOS1+MOS and pn in the second observation where \at~was the brightest.

The bottom panel of Fig.~\ref{fig:om_epic_lightcurves} shows the EPIC
light curves of \at~in the 0.5--7.3\,keV energy range. 
Within each 5\,h-long observations, \at~exhibited a nearly
constant X-ray flux. The minimum X-ray flux, that we will call hereafter
the quiescent level, was observed during the observation \#5, where
the averaged EPIC count rate was $0.030\pm0.002$\cnts. On a
timescale of days to weeks, the X-ray flux varies by a factor of 2--8,
except between the end of the first observation and the beginning of
the second observation, where the EPIC
count rate jumped in less than 2 days from $0.093\pm0.003$ to
$1.42\pm0.01$\cnts, i.e.\ a level $47\pm3$ times stronger than the
quiescent level. Then, the EPIC count rate decayed in less than 2 days
to $0.075\pm0.003$\cnts, i.e.\ a level only $2.5\pm0.2$ times
stronger than the quiescent level.

Such large amplitudes in the X-ray fluxes of young stellar objects are
usually observed during X-ray flares, which have typical light
curves with fast rise and peak phase, and slower (exponential)
decay phase, associated with fast heating and slow cooling of the
magnetically confined plasma
\citep[e.g.,][]{imanishi03,favata05}. X-ray flares with unusually long
rise phases have also been reported
\citep[e.g.,][]{grosso04,wang07,broos07}.

\onlfig{2}{
\begin{figure*}[!h]
\centering
\includegraphics[angle=0,width=2\columnwidth]{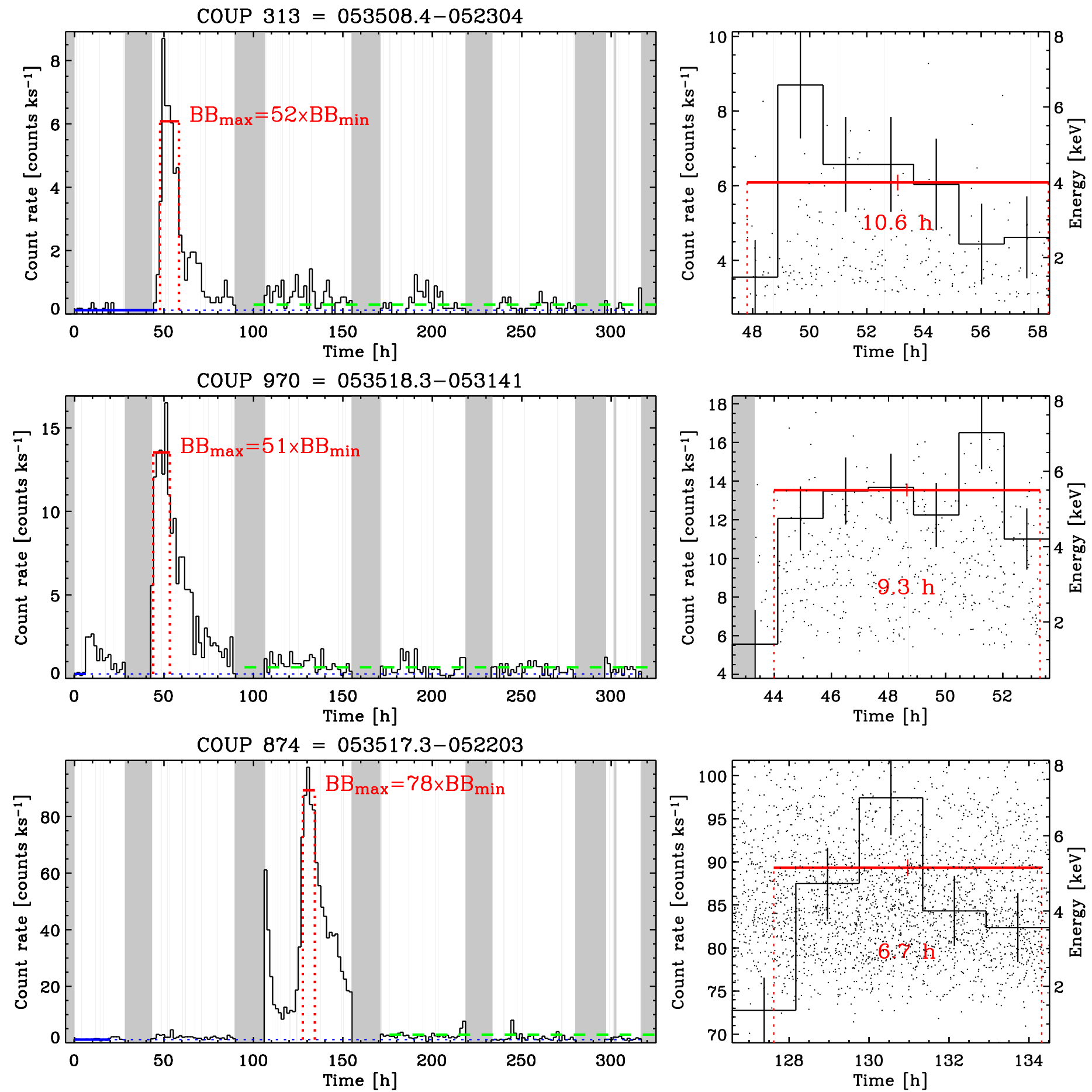}
 \caption{A subset of X-ray flares from the {\sl \cxo~Orion
     Ultradeep Project} with peak amplitude and duration larger than
     the one observed in \at. The left panels show COUP light curves
     \citep[see ][]{getman05b}, where blue and red segments
     indicate the minimum ($BB_{\rm min}$) and maximum ($BB_{\rm
     max}$) levels obtained from Bayesian block analysis
     \citep{scargle98}, respectively. The blue dotted line show the
     minimum level. The peak amplitudes are also given. The dashed green
     lines start 1.7 days after the end of the maximum Bayesian block,
     and indicate the level (2.5 times above the minimum level), that
     was observed in the third observation of \at. The right panels
     are an enlargement of the light curve around the Bayesian block
     showing the peak level and duration. Dots mark the arrival times of
     individual X-ray photons with their corresponding energies given
     on the right-hand axis. Large vertical gray stripes indicate the
     five passages of \cxo~through the van Allen belts where ACIS was
     taken out of the focal plane and thus was not observing Orion.
}
\label{fig:coup_lc}
\end{figure*}
}

\onlfig{3}{
\begin{figure*}
\centering
\begin{tabular}{@{}cc@{}}
\includegraphics[angle=0,width=0.9\columnwidth]{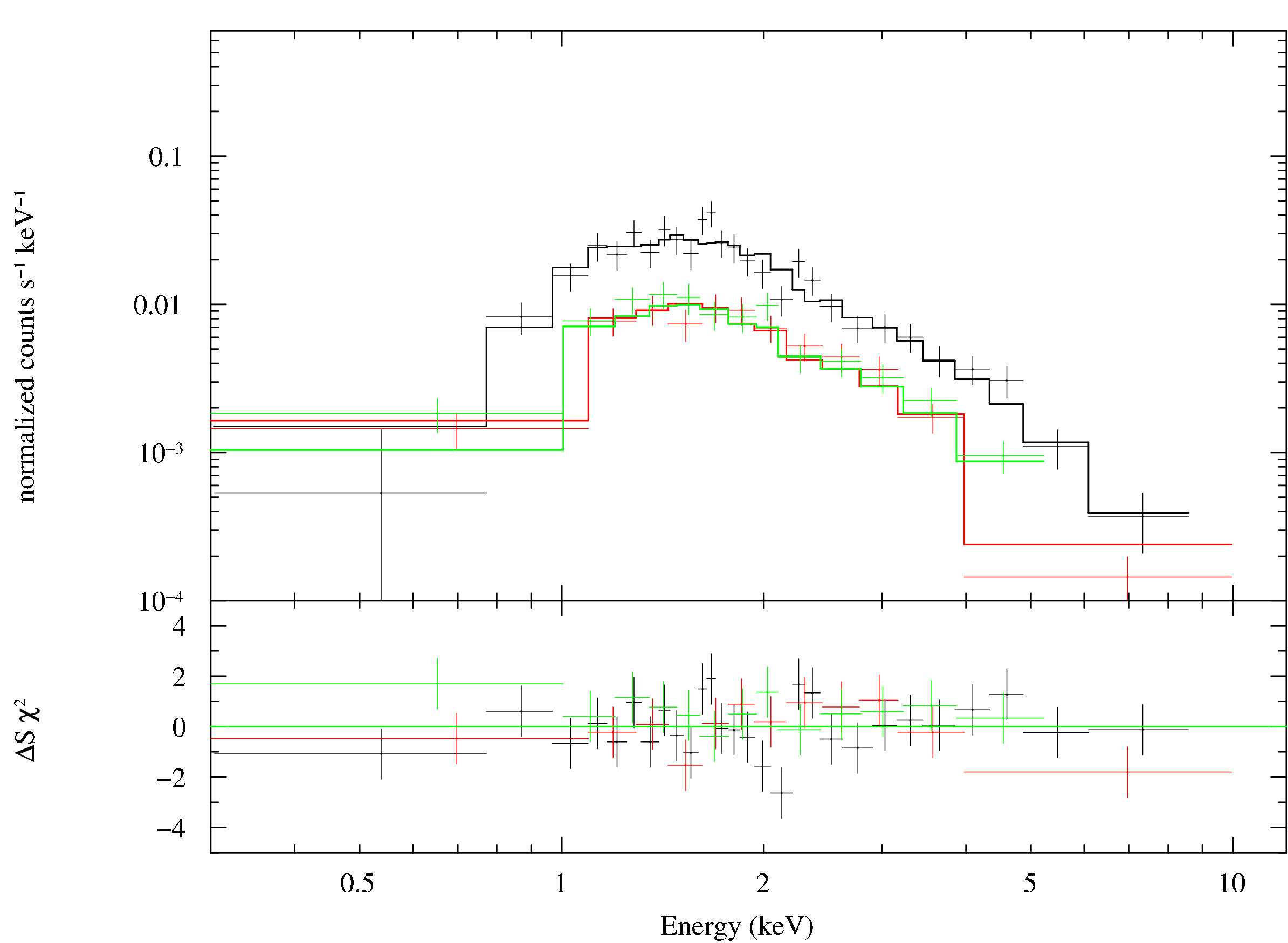}
& \includegraphics[angle=0,width=0.9\columnwidth]{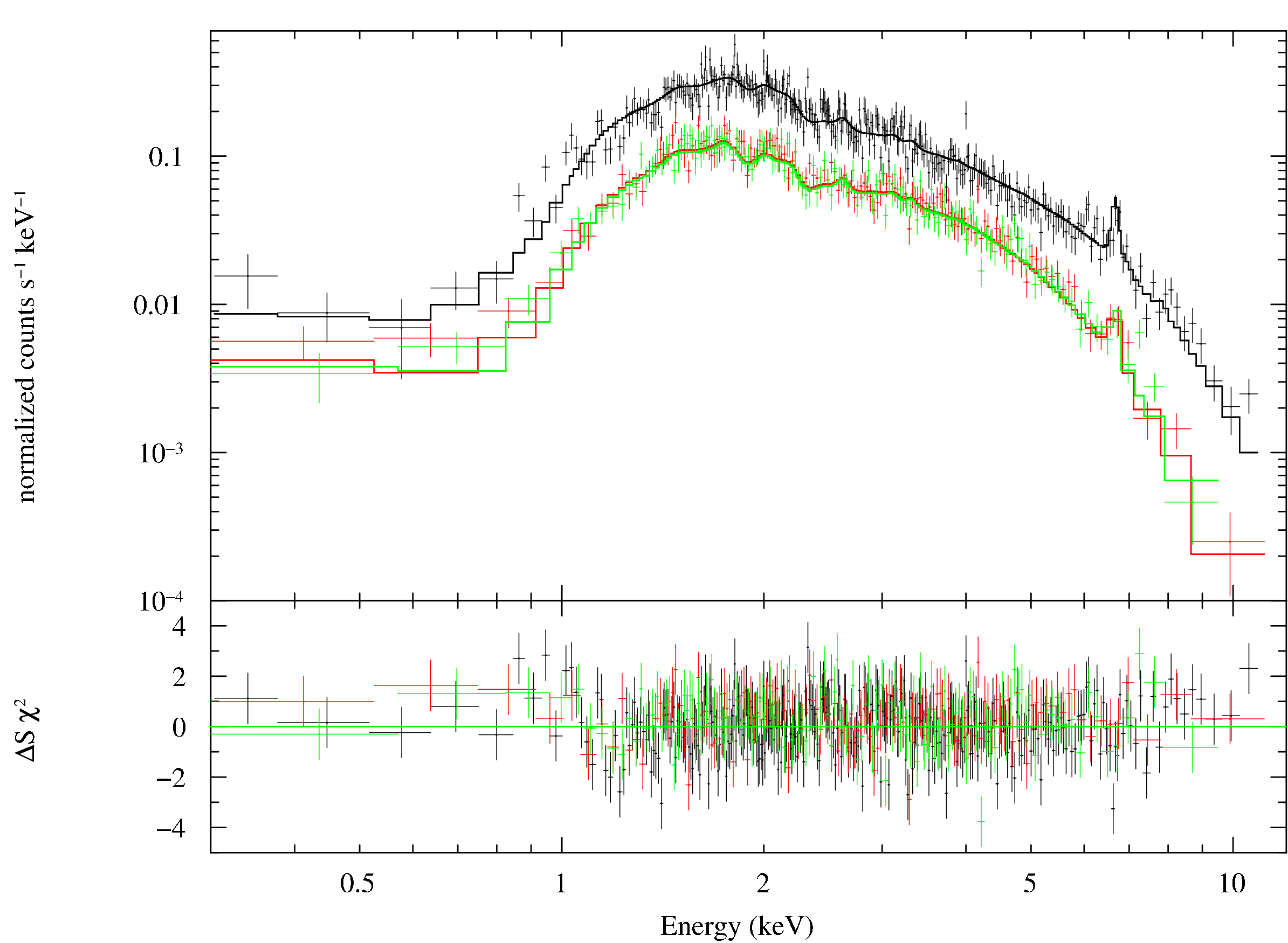} \\
\includegraphics[angle=0,width=0.9\columnwidth]{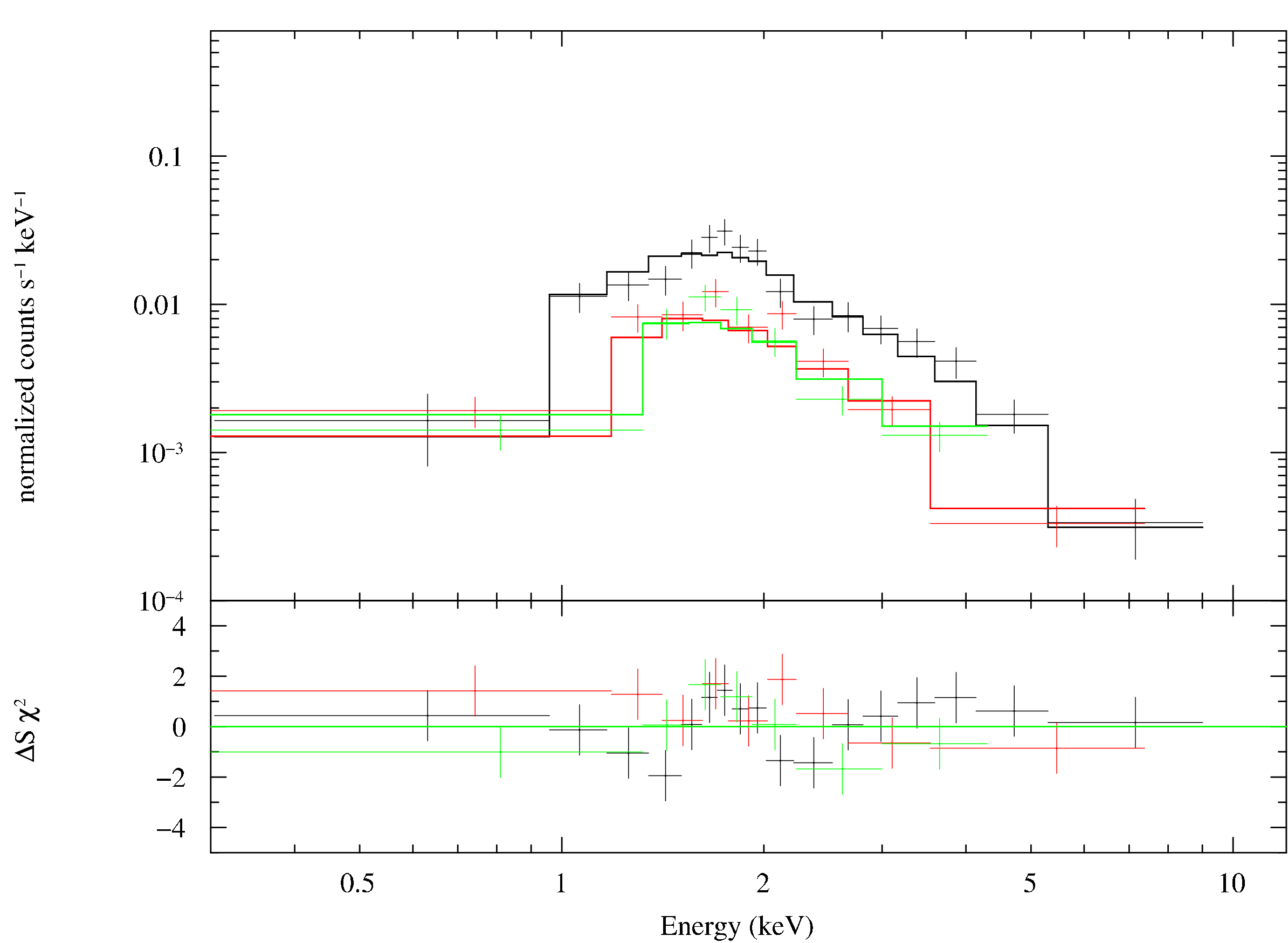} & 
\includegraphics[angle=0,width=0.9\columnwidth]{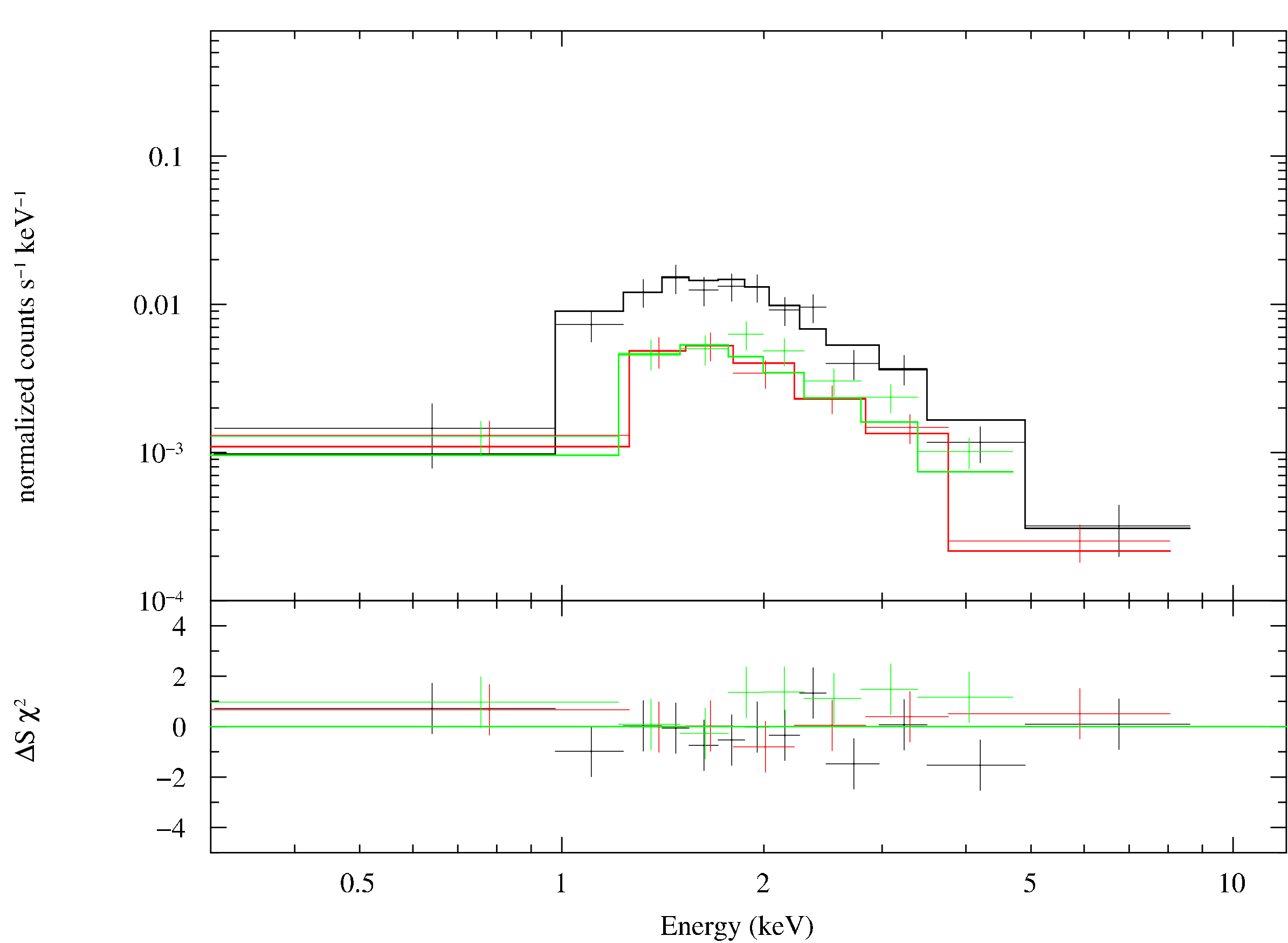} \\
\includegraphics[angle=0,width=0.9\columnwidth]{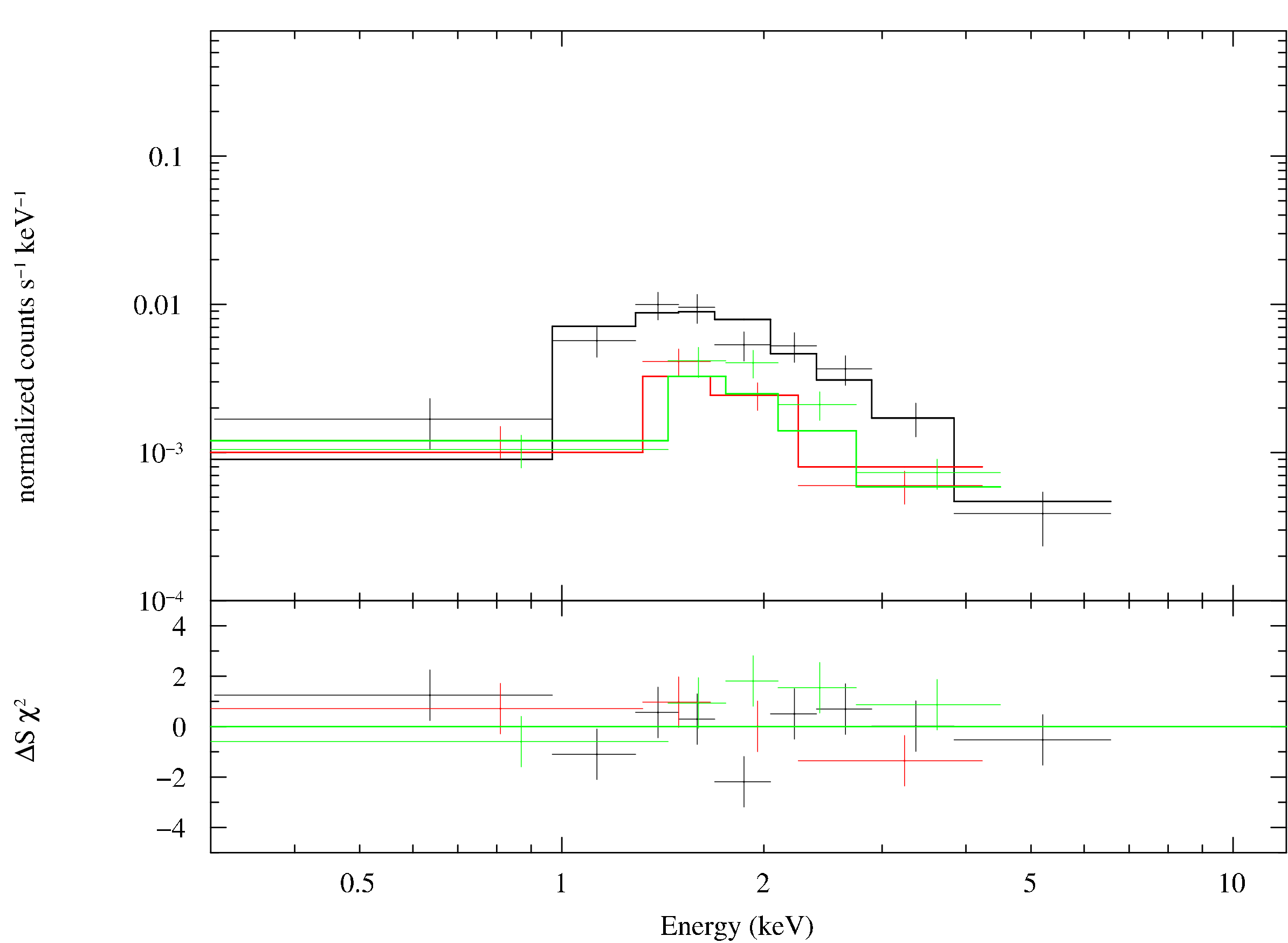} & 
\includegraphics[angle=0,width=0.9\columnwidth]{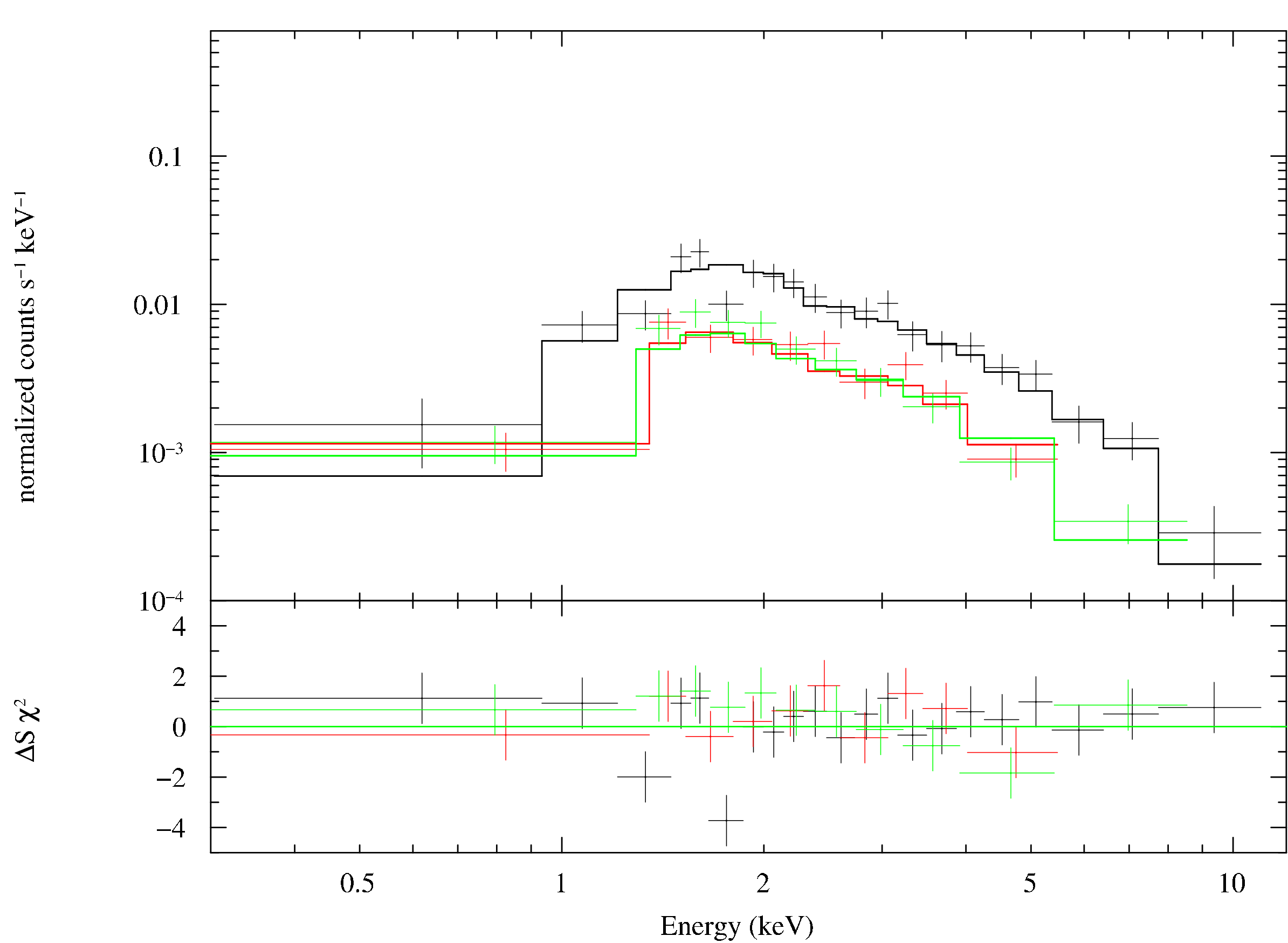}
\\
\includegraphics[angle=0,width=0.9\columnwidth]{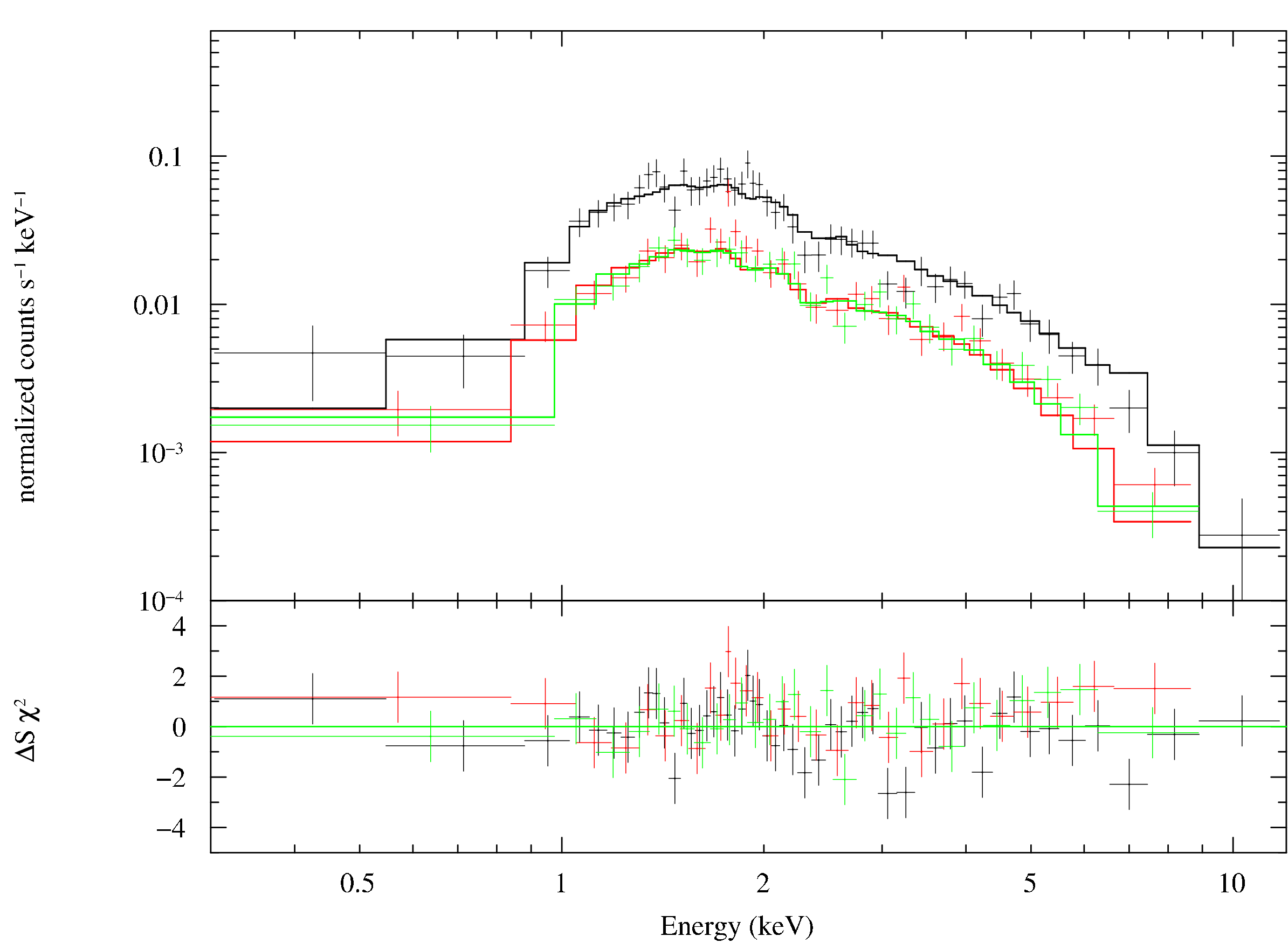} & 
\includegraphics[angle=0,width=0.9\columnwidth]{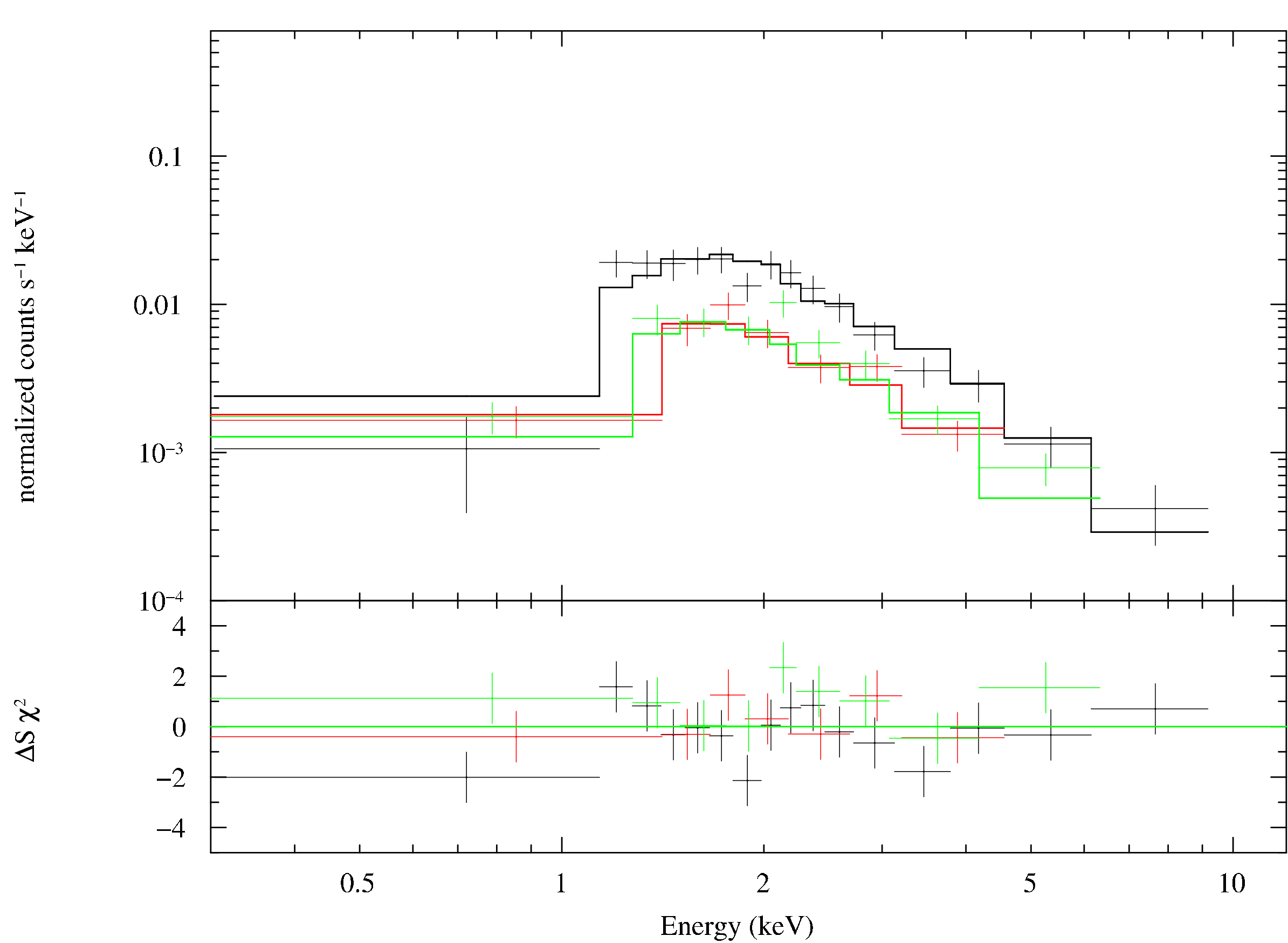}
%%\includegraphics[angle=-90,width=0.45\columnwidth]{epic_wabs_vapec_angr_xest_spectrum_obs1.ps}
%%& \includegraphics[angle=-90,width=0.45\columnwidth]{epic_wabs_vapec_angr_xest_spectrum_obs2.ps} \\
%%\includegraphics[angle=-90,width=0.45\columnwidth]{epic_wabs_vapec_angr_xest_spectrum_obs3.ps} & 
%%\includegraphics[angle=-90,width=0.45\columnwidth]{epic_wabs_vapec_angr_xest_spectrum_obs4.ps} \\
%%\includegraphics[angle=-90,width=0.45\columnwidth]{epic_wabs_vapec_angr_xest_spectrum_obs5.ps} & 
%%\includegraphics[angle=-90,width=0.45\columnwidth]{epic_wabs_vapec_angr_xest_spectrum_obs6.ps} \\
%%\includegraphics[angle=-90,width=0.45\columnwidth]{epic_wabs_vapec_angr_xest_spectrum_obs7.ps} & 
%%\includegraphics[angle=-90,width=0.45\columnwidth]{epic_wabs_vapec_angr_xest_spectrum_obs8.ps}
\end{tabular}
 \caption{EPIC pn (black), MOS1 (red), MOS2 (green) spectra of
 \at~plotted with the same scale. The lines show our best fits using
 one-temperature plasma combined with photoelectric absorption for
 observations \#1 to \#8
 (Table~\ref{table:fit_epic_wabs_vapec_angr_xest_parameters}) from
 left to right and top to bottom. The residuals are plotted in terms
 of sigmas with error bars of size one.
}
\label{fig:spectra}
\end{figure*}
}

\begin{table*}[!ht]
\caption{Best parameters of simultaneous fitting of EPIC pn, MOS1,
  MOS2 spectra with {\tt XSPEC}.}
\label{table:fit_epic_wabs_vapec_angr_xest_parameters}
\begin{tabular}{@{}c@{}ccccccrrrrcrc@{}}
\hline
\hline
\noalign{\smallskip}
                        &  
& &  \multicolumn{2}{c}{Plasma temperature} &
\multicolumn{2}{c}{Emission measure}   &
\multicolumn{2}{c}{Goodness-of-fit} & Flux$^\mathrm{\,b}$ &
\multicolumn{3}{c}{Intrinsic luminosity} & $\eta^\mathrm{\,c}$ \\
\vspace{-0.6cm}\\
                          &                    &     &
                          \multicolumn{2}{c}{\hrulefill} &   \multicolumn{2}{c}{\hrulefill} &
                          \multicolumn{2}{c}{\hrulefill}  &  & \multicolumn{3}{c}{\hrulefill} & \\
Obs. & \multicolumn{1}{c}{Rate$^\mathrm{\,a}$} & $N_\mathrm{H}$&
                          \multicolumn{1}{c}{$T_1$} &
                          \multicolumn{1}{c}{$T_2$} &
                          \multicolumn{1}{c}{$EM_1$} &
                          \multicolumn{1}{c}{$EM_2$}
                          & \multicolumn{1}{c}{$\chi^2_\nu$ ($\nu$)} &
                          \multicolumn{1}{c}{$\cal{Q}$} &  &
			  \multicolumn{3}{c}{0.5--2, 2--8, 0.5--8\,keV} \\
                & \multicolumn{1}{c}{(cnts\,s$^{-1}$)} & ($10^{22}$\,cm$^{-2}$)   & \multicolumn{2}{c}{(MK)}  & \multicolumn{2}{c}{($10^{53}$\,cm$^{-3}$)} & & \multicolumn{1}{c}{(\%)} &   & \multicolumn{3}{c}{($10^{30}$\,erg\,s$^{-1}$)}    \\
\noalign{\smallskip}
\hline
\noalign{\smallskip}
1 & 0.057  & 0.95 & \dotfill & 25.9 & \dotfill &  1.0 & 0.93
(50) &  61 &  1.9 &  0.6 &  0.4 &  0.9 & -3.5 \\
 & $\pm$0.002 & 0.87--1.02 & \dotfill & 23.9--28.4 & \dotfill &  0.9--1.0 \\
2 & 0.800  & 1.54 & \dotfill & 43.9 & \dotfill & 14.0 & 1.09
(599) &   5 & 37.8 &  8.2 &  9.2 & 17.4 & -2.2 \\
 & $\pm$0.008 & 1.51--1.57 & \dotfill & 42.7--45.2 & \dotfill & 13.6--14.3 \\
2 & 0.800 & 1.80 & 14.6 & 59.4 &  9.8 &  9.3 & 1.03 (597) &  28 & 37.9 & 11.8 &  9.5 & 21.3 & -2.2 \\
 &$\pm$0.008 & 1.75--1.85 & 13.4--15.7 & 54.6--66.0&  8.8--10.7 &  7.8--10.8 \\
3 & 0.046 & 1.36 & \dotfill & 20.8 & \dotfill &  1.1 & 1.21 (30) &  20 &  1.6 &  0.7 &  0.3 &  1.0 & -3.5 \\
 &$\pm$0.002 & 1.23--1.51 & \dotfill & 18.4--23.4 & \dotfill &  1.0--1.3 \\
4 & 0.031 & 1.33 & \dotfill & 21.5 & \dotfill &  0.8 & 0.79 (25) &  76 &  1.1 &  0.5 &  0.2 &  0.7 & -3.6 \\
 & $\pm$0.002 & 1.17--1.52 & \dotfill & 18.7--24.8 & \dotfill &  0.6--0.9 \\
5 & 0.019 & 1.04 & \dotfill & 22.6 & \dotfill &  0.4 & 1.32 (15) &  18 &  0.7 &  0.2 &  0.1 &  0.4 & -3.9 \\
 & $\pm$0.002 & 0.85--1.27 & \dotfill & 18.3--28.8 & \dotfill &  0.3--0.5 \\
6 & 0.050  & 1.54 & \dotfill & 40.5 & \dotfill &  0.9 & 1.20
(39) &  18 &  2.3 &  0.5 &  0.5 &  1.1 & -3.5 \\
 &$\pm$0.002 & 1.40--1.72 & \dotfill & 34.6--47.4 & \dotfill &  0.8--1.0 \\
7 & 0.147 & 1.02 & \dotfill & 44.8 & \dotfill &  2.1 & 1.05 (110) &  33 &  6.5 &  1.2 &  1.4 &  2.6 & -3.1 \\
 &$\pm$0.005 & 0.97--1.08 & \dotfill & 41.3--48.6 & \dotfill &  2.0--2.2 \\
8 & 0.043 & 1.53 & \dotfill & 23.3 & \dotfill &  1.2 & 1.19 (29) &  23 &  1.8 &  0.7 &  0.4 &  1.1 & -3.4 \\
 &$\pm$0.002 & 1.39--1.69 & \dotfill & 20.7--26.5 & \dotfill &  1.0--1.4 \\

\hline
\end{tabular}
Notes: We fit the X-ray spectra with the
  continuum and emission lines produced by an optically thin plasma in
  thermal collisional ionization equilibrium model ({\tt vapec}). We
  use for the plasma element abundances typical values observed in the
  coronae of young stars with fine X-ray spectroscopy \citep[][see
  Table~\ref{table:xest_angr_aspl_abund}]{guedel07}. The {\tt wabs}
  photoelectric absorption model use photoionization cross sections of
  \citet{morrison83}, and solar abundances of \citet{anders89}. Errors
  are given at the 68\% confidence level (i.e.\ $\Delta\chi^2=1.0$ for
  each parameter of interest),
  that corresponds to $1\sigma$ for Gaussian statistics. For
  observation \#2, a better fit is obtained using a plasma with two
  temperature components.
\begin{list}{}{}
\item[$^{\mathrm{a}}$] pn count rate (0.2--12\,keV).
\item[$^{\mathrm{b}}$] Observed X-ray flux (0.5--8.0\,keV) in unit of $10^{-13}$\,erg\,cm$^{-2}$\,s$^{-1}$.
\item[$^{\mathrm{c}}$] Logarithm of the X-ray intrinsic luminosity
  (0.5--8\,keV) to the bolometric luminosity ($0.8$\,L$_\odot$) ratio.
\end{list}
\end{table*}

For a comparison purpose, we note that, assuming a
typical convertion ratio between \xmm/EPIC and \cxo/ACIS-I count rates
\citep[e.g.,][]{ozawa05}, the average EPIC count rate of \at,
$\sim$0.08\cnts~at the distance of the Taurus molecular cloud
($d\sim140$\,pc), would convert to $\sim$0.001~\cxo/ACIS-I\cnts at
the distance of the Orion nebula ($d\sim450$\,pc). \at, put in the
Orion nebula cluster, would have then been brighter than 67\% of the X-ray
sources in the COUP. To check whether the behaviour of the light curve
of \at~is consistent with the one of an X-ray flare, we can compare it
to the light curve data set obtained in the COUP. 

We use the COUP results of the Bayesian block (BB)
variability analysis (developed by \citealp{scargle98}; adapted for
the COUP data set and coded in {\tt IDL} by one of us, N.G.), which segmented
the X-ray light curves into a contiguous sequences of constant count
rates \citep[see ][]{getman05b}. We define a source to be variable, if
there is more than one BB; with $BB_{\rm min}$ and $BB_{\rm max}$, the
minimum and the
maximum count rate levels, respectively (see, e.g.,
\citealp{stassun07}, for an application of COUP time-averaged X-ray
variability). The latter and the former are viewed as the quiescent
level and the peak level of the brightest flare,
respectively. Applying this criteria, there are 977 variable sources
(out of 1616 COUP sources). We find only 20 variable sources with
peak amplitude ($BB_{\rm max}/BB_{\rm min}$) and duration larger
than 45 and 4.7\,h., respectively, as observed for \at. For
comparison, our subsample sources have $BB_{\rm
  max}=10^{-4}$--2\cnts, i.e., they are at peak between 200 times
fainter and 100 times brighter than \at~(at the distance of the Orion
nebula) at peak.

Then, we make a visual examination of the selected BB
light curves to eliminate sources with peak flare BB having a spurious
long duration (including for example a passage through the van Allen
belts), and/or sources with a decay phase too slow to reproduce the
level observed at the beginning of the third observation of \at. For an
exponential decay phase with a decay timescale $\tau_{\rm d}$, this
latter criteria is equivalent to $\tau_{\rm d} < 0.6$\,day. Finally,
we exhibit the brightest flares from COUP 313, 874, and 970, which
fullfill our criteria (see online Fig.~\ref{fig:coup_lc}). These
sources are bona fine members of the Orion nebula cluster \citep{getman05a}. 

We conclude that a bright X-ray flare with a rapid cooling phase can
well reproduce the large amplitude, and also the flatness of the light
curve of \at~at its maximum. The following X-ray spectra analysis 
confirms this interpretation.

%________________________________________________________________
\subsection{X-ray spectra and plasma parameters}
\label{spectra}

\begin{figure}[!t]
\centering
\includegraphics[angle=0,width=0.9\columnwidth]{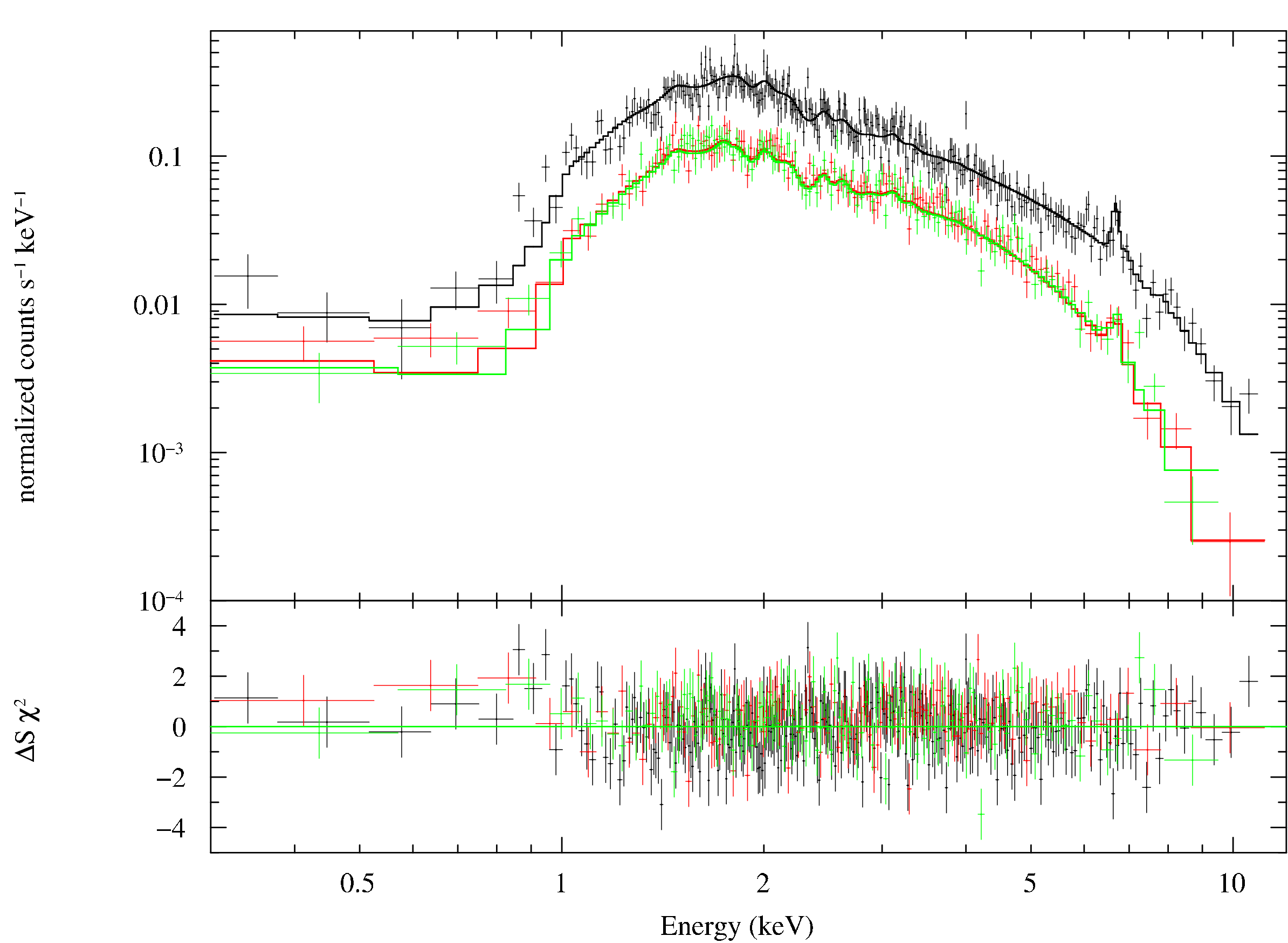}
 \caption{EPIC pn (black), MOS1 (red), MOS2 (green) spectra of \at~for
   observation \#2. The lines show our best fit using two temperature
   plasma combined with photoelectric absorption
   (Table~\ref{table:fit_epic_wabs_vapec_angr_xest_parameters}). The
   residuals are plotted in sigma units with error bars of size one.
}
\label{fig:obs2_spectra}
\end{figure}

For each observations the pn, MOS1, and MOS2
spectra were binned to 25 counts per spectral bin.
X-rays are detected up to 10\,keV (see online
  Fig.~\ref{fig:spectra}). The spectra are featureless, except in the
  second observation where a prominent line around 6.7\,keV is visible
  both in the pn and MOS spectra, corresponding to the Fe\,{\small
    XXV} triplet emission line (Fig.~\ref{fig:obs2_spectra}).

The pn, MOS1, and MOS2 spectra were fitted simultaneously
with {\tt XSPEC} \citep[version 12.3.0;][]{dorman01} to derive the
plasma parameters. The model is an X-ray emission spectrum from
collisionally-ionized diffuse gas, output from the Astrophysical
Plasma Emission Code ({\tt vapec}\footnote{More information can be
  found at:
  \href{http://hea-www.harvard.edu/APEC/sources_apec.html}{http://hea-www.harvard.edu/\-APEC/sources\_apec.html}\,.}), 
that includes continuum and emission lines. The plasma element
abundances were fixed to typical values measured in the coronae of
young stars with grating X-ray spectroscopy (see online appendix
Table~\ref{table:xest_angr_aspl_abund}), to allow a direct comparison with
the {\sl \xmm~Extended Survey of the Taurus molecular cloud}
\citep[XEST;][]{guedel07}. This emission model is combined with the
      {\tt wabs} photoelectric absorption model, which is based on the
      photoionization cross sections of \citet{morrison83}, and the
      solar abundances of \citet{anders89}.
The online Fig.~\ref{fig:spectra} shows our best fits. For the second
observation, we find a better fit by adding a second temperature
component, which reduced the $\chi^2$ from  655.4 (for 599 degrees of
freedom; d.o.f.) to 617.2 (for 597 d.o.f.). Given the new and old
values of $\chi^2$ and number of degrees of freedom, a F-test indicates a
probability of $\sim$$10^{-8}$ for the null hypothesis; therefore, we
conclude that it is reasonnable to add this extra temperature
component to improve the fit. Our best fit with a two-temperature
plasma is shown in Fig.~\ref{fig:obs2_spectra}.

Table~\ref{table:fit_epic_wabs_vapec_angr_xest_parameters} gives the
corresponding plasma parameters.On timescale of 2-days, the
photoelectric absorption of the X-ray spectra is not constant, but the
observed relative variations are lower than a factor of two. The
minimum and maximum values of the corresponding column density are
$\sim$$1.0 \times 10^{22}$\,cm$^{-2}$ and $1.8 \times
10^{22}$\,cm$^{-2}$, respectively; the latter was observed during the
second observation\footnote{The increase of the spectrum slope
  between 1 and 2\,keV observed between observations \#1 and \#2, is
  not only due to the increase of the absorption, as argued in SR,
  but also for 20\% to the increase of the plasma temperature.}.
SR didn't specify the model of X-ray absorption that they used,
  but their one-temperature plasma model gave qualitatively similar
  results. We stress that the value of the column density can even be
increased by 50\% when revised solar abundances (i.e., metal poor) are
adopted for the abundances of the absorbing material (see online
Appendix~\ref{appendix:newabun}).

The plasma temperature is $\sim$23\,MK during the low activity
levels. The observations showing an increase of X-ray count rates
(namely \#2, \#6, and \#7) also correspond to phases with the highest
plasma temperatures (60\,MK, 40\,MK, and 45\,MK, respectively), and
therefore to flaring activity.  

During the second observation the X-ray surface flux, i.e., the
intrinsic X-ray luminosity divided by the stellar surface, 
peaked to $10^8$\ergscm. The temperatures of the hot (60\,MK) and cool (15\,MK) plasma
components, associated with this elevated level of X-ray surface flux,
are consistent with the ones observed in the most active TTSs of COUP
\citep[see Fig.~11 of][]{preibisch05b}. The cool plasma component is
usually observed in the coronae of active stars, and may define a fundamental
coronal structure, which is probably related to a class of compact
loops with high plasma density \citep[see][ and references therein]{preibisch05b}.

We make an estimate of the hot loop length using the method of
\citet{reale97} \citep[see also][for an application on COUP
  data]{favata05}. Assuming a peak temperature of 60\,MK and a flare
decay time lower than 0.6 day (see Sect.~\ref{lc}), we find for the
semi-circular loop a half-length lower than 7.1\,R$_\star$ for a
freely decaying loop, with no heating; or lower than 1.6\,R$_\star$
for a strongly sustained heating\footnote{\citet{reale97} use $\zeta$, the
  slope of the flare decay in a log-log diagram of the plasma
  temperature vs.\ the squared-root of the emission measure (a proxy
  of the plasma density), as diagnostic to assess the level of
  sustained heating in the analysis of stellar flares. For the
  \xmm~energy coverture, the corresponding range for $\zeta$ values are 0.4
  and 1.9, for strongly sustained heating and freely decaying loop,
  respectively.} in the cooling phase.
Therefore, the height of the semi-circular loop is lower than
1--4.5\,R$_\star$. Therefore, this loop cannot connect the
stellar surface and the inner accretion disk, distant by 7.8 stellar
radii, and is likely anchored on the stellar photosphere.

The high temperature and X-ray surface flux observed during the second
observation point to an enhanced X-ray activity produced by a bright
X-ray flare. Such bright flares on the Sun are sometimes associated with
coronal mass ejection (CME). Therefore, we cannot rule out that the
maximum of column density observed during the second observation is
due to this energetic event.

%________________________________________________________________
\subsection{Comparison with previous X-ray observations of \at}
\label{archives}

\at~was previously observed several times in X-rays at different
epochs: on March 4, 1980 and
February 7, 1981 with the IPC on board {\sl Einstein}, with an exposure
of 0.6\,h and 2.8\,h, respectively; on August 10, 
1990 during the {\sl ROSAT All Sky Survey} (RASS) with the PSPC on
board {\sl ROSAT}, with an exposure of 0.2\,h; and on February 22 and
August 18, 1993 during two pointed PSPC observations of the young
binary Haro~6-13 (PI: H.\ Zinnecker), with
an exposure of 0.6\,h and 1.5\,h, respectively.

\citet{walter81} reported a detection with $0.030\pm0.005$\,IPC\cnts,
during the first observation with {\sl Einstein} (bright enough to
exhibit a crude spectrum), but only an upper
limit of 0.004\,IPC\cnts~with a 5 times longer observation, nearly one
year later \citep{walter84}.  \citet{walter84} associated the X-ray
emission detected from \at~as a quiescent level, and interpreted the
non-detection in the framework of the smothered coronae of TTSs
\citep[see][]{walter81}, where it was proposed that mass ejection
could increase enough the absorbing column density to smother the coronal
(quiescent) X-ray emission. \citet{neuhaeuser95} reported 10 years
later a RASS detection with $0.014\pm0.006$\,PSPC\cnts.\footnote{Note
  that SR reported a RASS rate 10 times higher than the one reported
  by \citet{neuhaeuser95}.} \at~was also
detected during the PSPC pointed observation with
$0.022\pm0.005$\,PSPC\cnts \citep[see in the WGA catalogue of {\sl
    ROSAT} point sources, the source \object{1WGA J0434.9+2428},
  located $37\arcmin$ off-axis;][]{white96}, but (again) only during
the shortest exposure.

Further constraint can be derived on the X-ray variability of \at~by
combining these multiple epoch observations with our better assessment
of its quiescent level thanks to better spectra. For a plasma
temperature $\sim$23\,MK and an unabsorbed X-ray luminosity in the
energy band from 0.5 to 8.0\,keV of
$\sim0.5\times10^{30}$\,erg\,s$^{-1}$ (equivalent to an unabsorbed
X-ray flux of $2.1\times 10^{-13}$\ergscm), absorbed by a column
density of $\sim10^{22}$\,cm$^{-2}$, we compute with {\tt
  PIMMS}\footnote{Available at
  \href{http://asc.harvard.edu/toolkit/pimms.jsp}{\tt
    http://asc.harvard.edu/toolkit/pimms.jsp}\,.} that {\sl
  Einstein}/IPC (0.2--4.5\,keV) and {\sl ROSAT}/PSPC (0.12--2.48\,keV)
would both observe only $\sim$0.002\cnts. This low count rate is
consistent with the IPC upper limit,\footnote{However, the
  RASS rate is not consistent with the IPC upper limit. Note that SR
  argued the opposite.} and we found that it is twice lower than the
background level (inside a 1.5\arcmin-radius circle) at
the location of \at~in the second PSPC pointed observation. Therefore,
the previous non-detections in X-rays are consistent with the
quiescent level observed with \xmm.

We conclude that the previous X-ray detections with {\sl Einstein},
the RASS, and {\sl ROSAT/PSPC}, were made during high levels of
activity, where \at~was about 18, 8, and 13 times, respectively, above
its quiescent level.

%________________________________________________________________
\subsection{\xmm~optical/UV monitor light curves}
\label{om}

\subsubsection{UV photometry}

The OM was operated in the
imaging mode default, which uses 5 imaging windows plus a small
(1.7\arcmin$\times$1.7\arcmin) central imaging window. The imaging
mode default consists of a sequence of 5 exposures where one of the 5
imaging windows covers a large fraction of the OM field-of-view
(17\arcmin$\times$17\arcmin) with $1\arcsec\times1\arcsec$ spatial
resolution, while the small central imaging window ensures a
continuous monitoring of the target at the center of the field-of-view with
$0\farcs5\times0\farcs5$ spatial resolution (for an illustration
  of this exposure sequence and a light curve obtained with the
  small central window see 
  \href{http://xmm.vilspa.esa.es/external/xmm_user_support/documentation/uhb/node69.html\#uhb:fig:nicerudi5}{Fig.\ 85} of the {\sl \xmm~Users' Handbook} and \citealt{grosso07b}, respectively). We requested the minimum available exposure time in imaging mode default (800\,s) to monitor any change in the UV photometry (UVW2 filter at 206\,nm) with a $\sim$15\,min sampling rate.

We run the OM imaging mode pipeline. We made our own {\tt IDL} program to
plot the light curves of \at~from the source lists of the
small central window (*OM*SWSRLI0000.FIT), and the large central window
(*OM*SWSRLI1000.FIT). The latter provides {\sl only} 1 data point per
OM exposure sequence (about $5\times800$\,s), which is obtained
simultaneously with the first (out of 5) data point of the small
central window. In a few exposures, \at~is missing in the observation
source list of the small central window, but a visual inspection of
the corresponding images confirms the detection. Therefore, we
complete the light curve by doing aperture photometry. 

The top panel of Fig.~\ref{fig:om_epic_lightcurves} shows the UV light
curves of \at. We note that the UV light curves of \at~reported in
SR were limited to the UV photometry obtained with the
large OM central imaging window (see for comparison the grey
hourglasses in the top panel of our Fig.~\ref{fig:om_epic_lightcurves}
and their Fig.~1). The continuous monitoring with the central window
is crucial to determine accurately the UV variations on hour
timescale. We find that \at~is variable in UV by an amount of a few
0.1\,mag on a few hours timescale, and up to 1\,mag on a week
timescale. There are no correlation between the UV and X-ray variations.

\begin{figure}[!t]
\centering
\includegraphics[angle=0,width=\columnwidth]{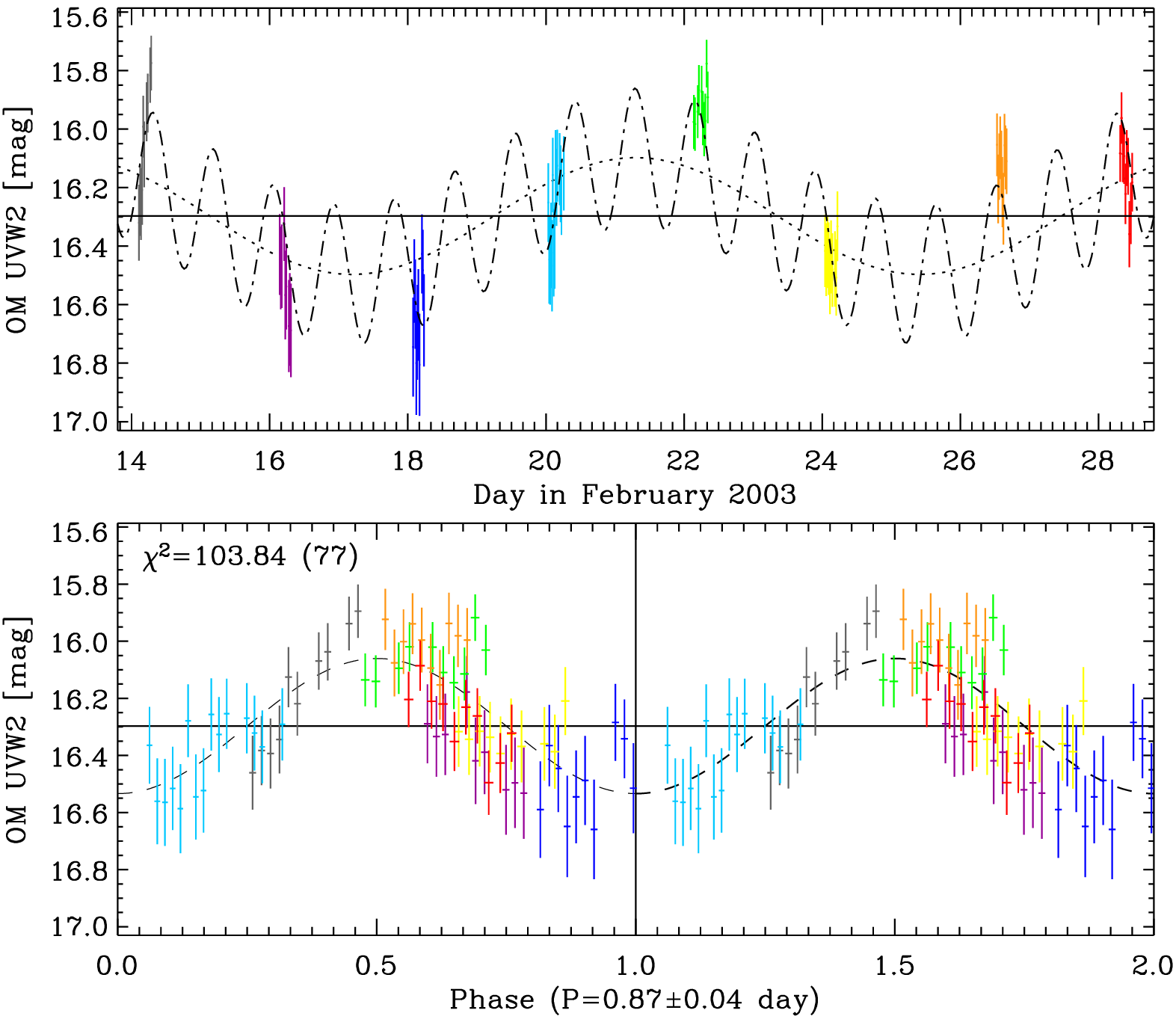}
 \caption{Weekly and daily UV variability of
 \at~observed with the OM. Top panel: the dotted and dashed-dotted
 line show the weekly modulation of the UV flux attributed to the
 eclipse period (8.22 days), and the overall modulation of the UV
 flux. The horizontal line shows the average flux. Bottom panel: the
 dashed line shows the daily modulation of the UV folded in phase
 after subtraction of the weekly modulation. Symbols are the same in
 both panels. The formula of the overall fit (dashed-dotted line) is
 given by Eq.~\ref{fit_uv}.}
\label{fig:om_period}
\end{figure}

\subsubsection{Modeling of the UV variability}

The UV excess observed in CTTSs is attributed to the accretion shocks
at the base of the accretion funnel flows \citep[see review on
  magnetospheric accretion by][]{bouvier07b}. Therefore, we assume
that the UV flux must be somehow modulated by the warped inner disk,
and we perform a least squares fitting of the UV light curve using a
cosine function with free amplitude and phase, and period fixed to
8.22 days \citep{bouvier07}. The resulting fit ($\chi^2=206.5$ for 80
d.o.f.) is not acceptable, mainly due to fast
variations on day timescale that cannot be properly reproduced
with this simple model including only weekly variations. 
A Lomb periodogram of the residuals suggests the presence of an additional
(high-frequency) modulation with a period of $0.87\pm0.03$ day. 

We then compute a grid of fits in the amplitude and phase parameter
space to look for optimized values that maximize the peak of the
periodogram residuals. This is equivalent to a simultaneous fitting of
both modulations with the long-term period is fixed. We find a
better fit ($\chi^2=103.8$ for 77 d.o.f.). Given the new and old
values of $\chi^2$ and number of
degrees of freedom, a F-test indicates a probability of
$\sim$$10^{-11}$ for the null hypothesis; therefore, we conclude that
it is reasonnable to add this extra modulation to improve the
fit. Moreover, a Lomb periodogram of the residuals shows no other
modulations, and a perfect substraction of the two modulations. The
fit formula is given by the following equation:
\begin{eqnarray}
UV & = & 16.30 \pm 0.02\,{\rm mag}\label{fit_uv}\\
     & + & (0.20 \pm 0.02)\, \cos \{ 2\pi [t-(17.2 \pm  0.1)]/8.22 \} \nonumber\\
     & + & (0.24 \pm 0.02)\, \cos \{ 2\pi [t-(14.75 \pm 0.02)]/(0.87 \pm 0.04) \}  , \nonumber
\end{eqnarray}
where $t$ is the day in February
2003, and the errors are given at the 68\% confidence level.

The top panel of Fig.~\ref{fig:om_period} shows the UV light curve and
the fit given by Eq.~\ref{fit_uv}. The bottom panel shows the UV light
curve folded in phase with the high-frequency period after the
substraction of the long-term modulation. This light curve looks
  rather convincing. However, a longer UV observation with the
  OM is necessary to have a definitive confirmation of this
high-frequency period, which would suggest a non-steady accretion.

\begin{figure}[!t]
\centering
\includegraphics[angle=0,width=\columnwidth]{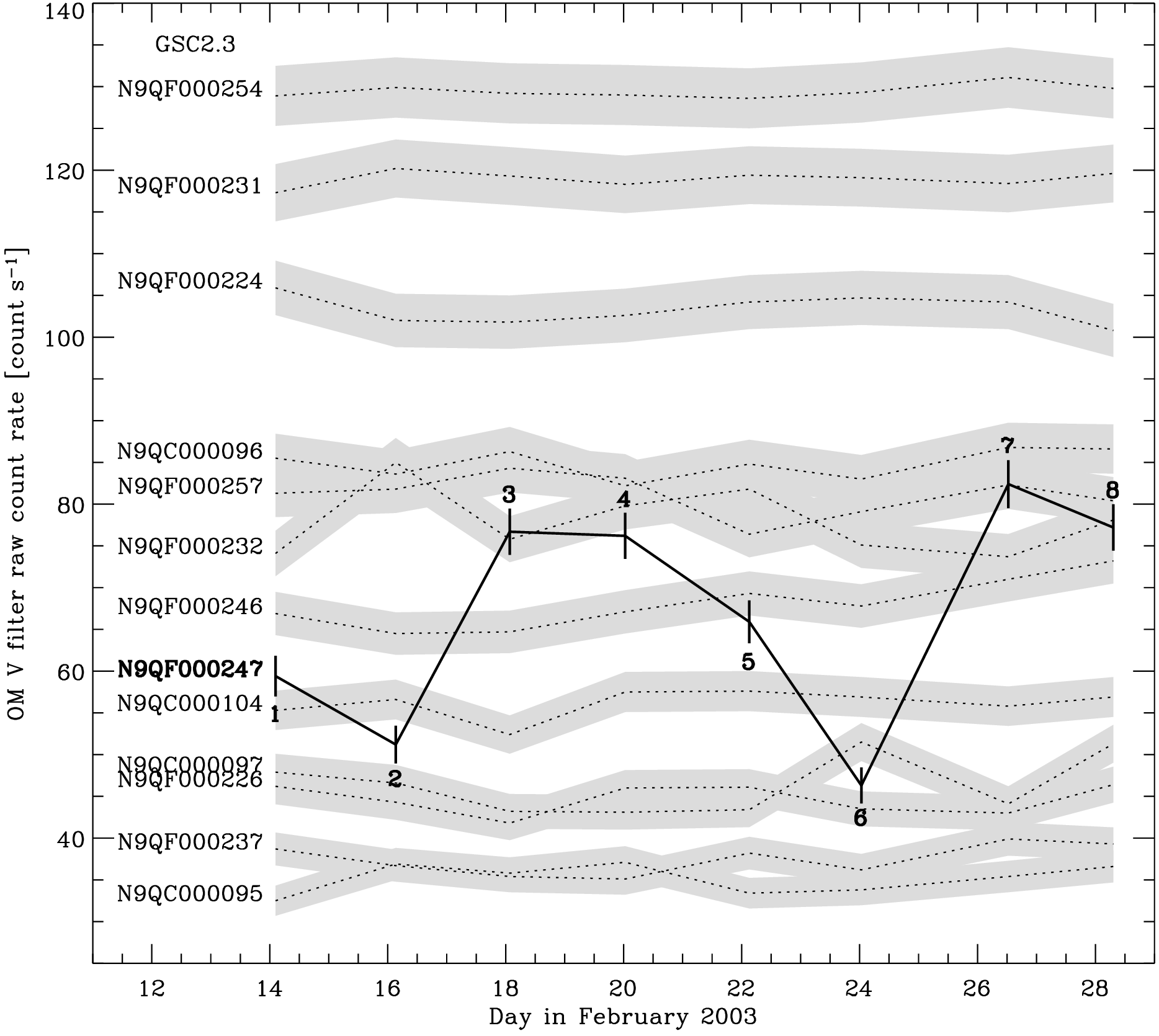}
 \caption{Optical light curves of the OM guide stars obtained with the
 field acquisition exposure. The dashed lines and grey stripes
 indicate the raw count rates of guide stars and one-sigma variations,
 respectively. The thick line and error bars shows the optical light
 curve of \at. Labels indicate counterparts in the Guide Star Catalog
 (version 2.3.2).
}
\label{fig:faq}
\end{figure}

\subsubsection{Optical variability from the field acquisition exposure}
\label{faq}

We propose here to obtain extra informations on the optical photometry
of the target thanks to the OM Field Acquisition exposures (FAQs),
where several stars are also detected in optical in the OM field of
view, and are then available as comparison stars for variability study.

The FAQ is a short exposure (10\,s) $V$-filter
image always taken at the start of each science observation (i.e., at
the beginning of the MOS observation) to allow proper identification
of guide stars and to compensate small pointing errors. A thresholding
is applied to the FAQ image by the OM on board software to identify
guide stars. The derived offset is then applied to the OM science
windows. The informations (positions in detector coordinates, count
rates,...) on the guide stars are reported in the {\tt *OMS40000RFX*}
data files delivered with the Observation Data Files (ODFs). 
The use of the FAQ image, instead of scheduling an additional OM
exposure in the imaging mode default with the $V$-filter, allowed us
to make a twice longer observation with the UVW2 filter, by saving at
least $5\times800$\,s (without taking into account time overheads).

In each observation, we recover sky coordinates from detector
coordinates, and identify the counterparts of guide stars in the
  Guide Star Catalog (version 2.3.2). Then, we build a light curve
for the 13 guide stars
(including \at). Fig.~\ref{fig:faq} shows the variations of the raw
count rates in the optical of the guide stars during our campaign. All
guide stars, except \at, exhibited small (lower than 2$\sigma$)
relative variations in the optical. By contrast, \at~exhibited large
relative variations, in particular, during observations \#1, \#2, \#5, and
\#6 where a large dip is visible. We conclude that these 4
\xmm~observations were likely made during the optical eclipses of \at.

\begin{figure*}[!ht]
\centering
\includegraphics[angle=0,width=1.82\columnwidth]{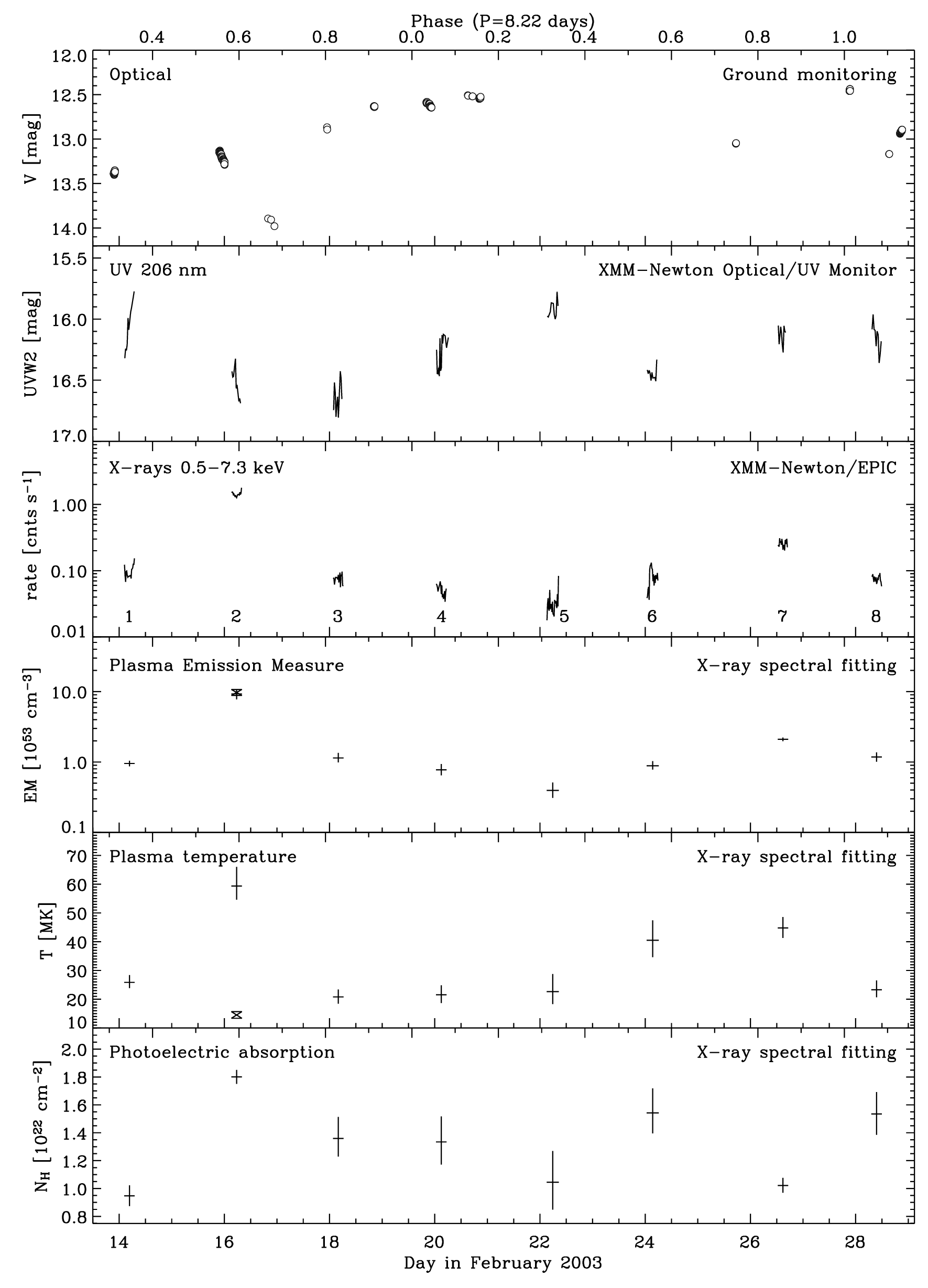}
 \caption{Light curves and plasma parameters of \at. The
   panels show from top to bottom: the optical, UV, and X-ray light
   curves (labels indicate observation numbers); the variations of the
   plasma emission measure(s) and temperature(s), and the
   photoelectric absorption
   (Table~\ref{table:fit_epic_wabs_vapec_angr_xest_parameters}). The
   top horizontal axis indicates the corresponding phase for the
   rotation period of 8.22 days. The (arbitrary) phase origin is from
   \citet{bouvier07}.
}
\label{fig:lightcurves}
\end{figure*}

\begin{figure*}[!ht]
\centering
\includegraphics[angle=0,width=1.82\columnwidth]{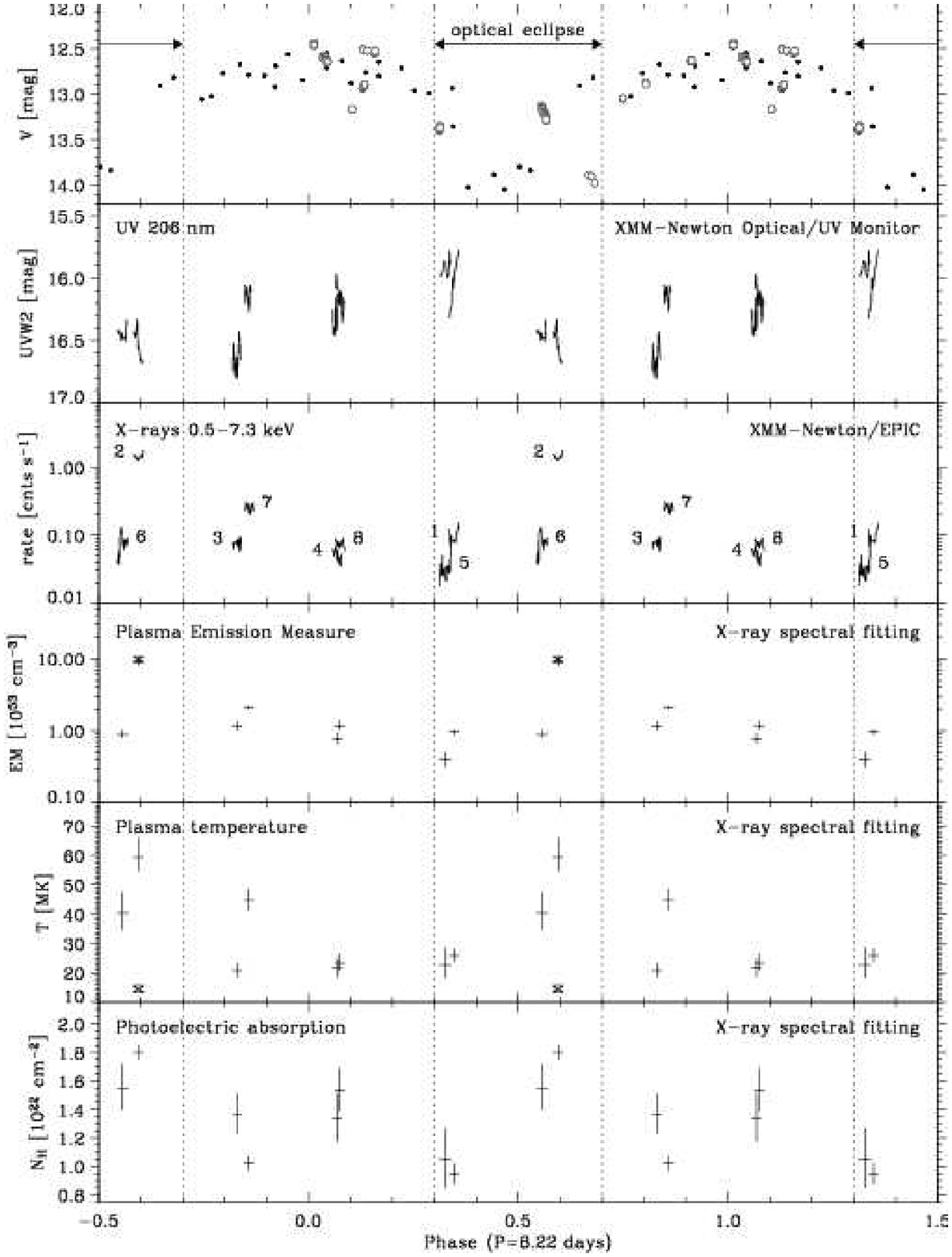}
 \caption{Light curves and plasma parameters of \at~ folded in phase
   with the rotation period of 8.22 days. Symbols and the phase origin are
   identical to the ones used in Fig.~\ref{fig:lightcurves}. In the top
   panel, black dots show for comparison the optical ground monitoring
   from Aug.\ 27 to Oct.\ 24, 2003 \citep{bouvier07}. The horizontal
   arrow shows our estimate of the eclipse phase based on the optical
   ground monitoring.
}
\label{fig:folded_lightcurves}
\end{figure*}
%________________________________________________________________
%________________________________________________________________
\section{UV and X-ray variabilities versus rotational phase}
\label{variability}

\subsection{Ground-based optical photometry}

The Journal of ground-based optical observations is given in
Table~\ref{log}. Observations were carried out from three sites over a
time span of two weeks, covering our \xmm~observations in February
2003, using either CCD detectors or a photomultiplier
tube (Mt Maidanak). Measurements were obtained in the $V$
filter. Differential
photometry was performed on CCD images and absolute photometry from
photomultiplier observations, with an accuracy of the order of
0.01\,mag. Somewhat larger systematic errors
($\leq$0.05\,mag) might result from the relative calibration of the
photometry between sites. All data reduction procedures can be found
in \citet{bouvier03}.

The top panel of Fig.~\ref{fig:lightcurves} shows the ground-based
optical light curve; for comparison purpose, the other panels show the UV
and X-ray light curves, and the plasma parameters. Unfortunately, due
to bad weather conditions in Europe, none ground telescope was able to
obtain simultaneous optical observations with \xmm. In the most
favorable cases, ground optical photometry was obtained only 4.3\,h,
3.1\,h, 2.8\,h, and 2.0\,h before the beginning of the MOS
observations \#1--\#4, respectively; and only 3.3\,h after the end the
last MOS observation. Despite the limited time sampling, a dip, likely
associated with a primary eclipse of \at, is however detected at the
beginning of the monitoring campaign. This supports what we deduced
previously in Sect.~\ref{faq} from the optical FAQ images, that
\xmm~observations \#1 and \#2 were obtained during the
eclipse. Consequently, the FAQ images can also be used safely to
identify an eclipse during \xmm~observations \#5 and \#6, located
within the Feb.\ 23--27, 2003 gap of the ground observation.
 
\begin{table}[t]
\caption{Journal of ground-based optical observations.}
\label{log}
\begin{tabular}{clclr}
\hline\hline
Feb.\ 2003           & \multicolumn{1}{c}{Site} & Tel. &Observer    &N$_{\rm obs}$\\  
(d) \\
\hline
13.89--28.88 & Teide (Spain)         & 0.8\,m  & M.R.\ Z.\ &177\\
16.83--17.96 & Sierra Nevada (Spain) & 1.5\,m  &  M.\ F.\  &5\\
20.63--28.64 & Mt Maidanak (Uzbek.)  & 0.5\,m  &  K.\ G.\  & 8\\
\hline
\end{tabular}
\end{table}

\subsection{Folded light curves and eclipse phases}

To obtain a better determination of the dates of the optical eclipses,
we compare our photometry with the one obtained just 6 months apart,
from Aug.\ 27 to Oct.\ 24, 2003 \citep{bouvier07}. The top panel of
Fig.~\ref{fig:folded_lightcurves} shows both photometry data set
folded together in phase using the 8.22-day rotational period
\citep{bouvier07}, i.e., we introduced no phase difference between
  the two data set. The (arbitrary) common origin of phase was
chosen to match the light curve of season 2003 shown in Fig.~15 of
\citet{bouvier07}, taking the eclipse center at phase 0.5. 

The behaviour of our photometry is consistent
with the shape of the light curve observed during the season 2003. We
can exclude any phase offset larger than 0.05 between the two epochs. 
A deep primary eclipse ($\Delta V\sim1.5$\,mag) is well detected at the
beginning of our monitoring campaign. There is also some evidence for
a shallow secondary eclipse detected only at the end of our monitoring
campaign, after the last \xmm~observation. 
The differences in brightness of about 0.5 and 1\,mag
between the Feb.\ and Aug.\ 2003 data, observed at the beginning and the
end of the primary eclipse, respectively, can be explained with a
longer duration of the eclipse in Feb.\ 2003.
The brightening on Feb.\ 16, 2003, close to the center of the primary
eclipse, was observed with a high sampling rate ($\sim2$\,min),
and exhibited only a fast decay. Therefore, this event is similar
to the transient brightening events usually observed in the faint
state of \at~\citep{bouvier99}.
The primary eclipses were centered on Feb.\ 15.5 and Feb.\ 23.5, 2003,
and covered phases ranging from $\sim$0.3 to $\sim$0.7. Our
ground-based monitoring confirms that \xmm~observations \#1, \#2, \#5,
and \#6 were made during primary eclipses. In particular, observations
\#2 and \#6 are secured close to the center of the primary eclipse.
We can exclude a secondary eclipse at the start of observations \#4 and
\#8 thanks to the FAQ, and moreover the corresponding X-ray light
curves show no decay. Therefore, we conclude that the shallow
secondary eclipse likely started after the end of our observations.

The UV flux is the highest when the primary eclipse starts and the
lowest towards the end of it. Indeed, the lowest UV flux was
observed during the \xmm~observation \#3 at the end of the egress
phase, i.e., outside the primary eclipse.  Our model of
the UV flux variations helps to disentangle weekly modulation (produced
 by the warped disk) and daily variation. Eq.~\ref{fit_uv} indicates
that the weekly modulation was at minimum on Feb. 17.2, 2003 (see also
the top panel fo Fig.~\ref{fig:om_period}), which corresponds to a phase
delay of about 0.2 compared to the optical eclipse. Therefore, the
warped disk produces a maximum of obscuration of the UV flux at the
end of the optical eclipse. Consequently, the true maximum of the UV flux
occurred around phase 0.2, well before the start of the eclipse. We didn't
survey this time interval with \xmm, but we note that a brightening in
the B-band of \at~was observed around phase 0.2 during the 1995
campaign \citep{bouvier99}. The delay between the optical and UV
eclipse, and the smaller depth of the UV eclipse ($\Delta UVW2\sim
0.40$\,mag) compared to the optical eclipse ($\Delta V\sim1.5$\,mag),
suggests a {\sl trailing} accretion funnel flow, producing a strong
absorption of the UV photons emitted at the accretion shock.

No eclipses are detected in X-rays. The variable photoelectric
absorption of the X-ray spectra is not correlated with the rotational
phase; similar low and high values of the column density are observed
both during the eclipse and outside it. The maximum of the column
density was observed at phase 0.6, close to the center of the optical
eclipse, during the X-ray flare. However, increases of the column
density were also reported during large stellar flares, and solar flares are
sometimes associated with coronal mass ejection \citep[see the review
  on X-ray astronomy of stellar coronae by][ and references
  therein]{guedel04b}. Therefore, we cannot rule out that the peak of
column density is due to this energetic event. 
The gas column density on the line of sight produced by the warped
disk is around $10^{25}$\,cm$^{-2}$ \citep{bouvier99}, which is large enough
to absorb all the X-rays emitted by \at, even during a bright and hot flare.
The lack of eclipses in X-rays indicates that X-ray emitting regions
are located above the high-density disk warp at high latitudes. 

\onlfig{9}{
\begin{figure*}[!h]
\centering
\includegraphics[angle=0,width=\columnwidth]{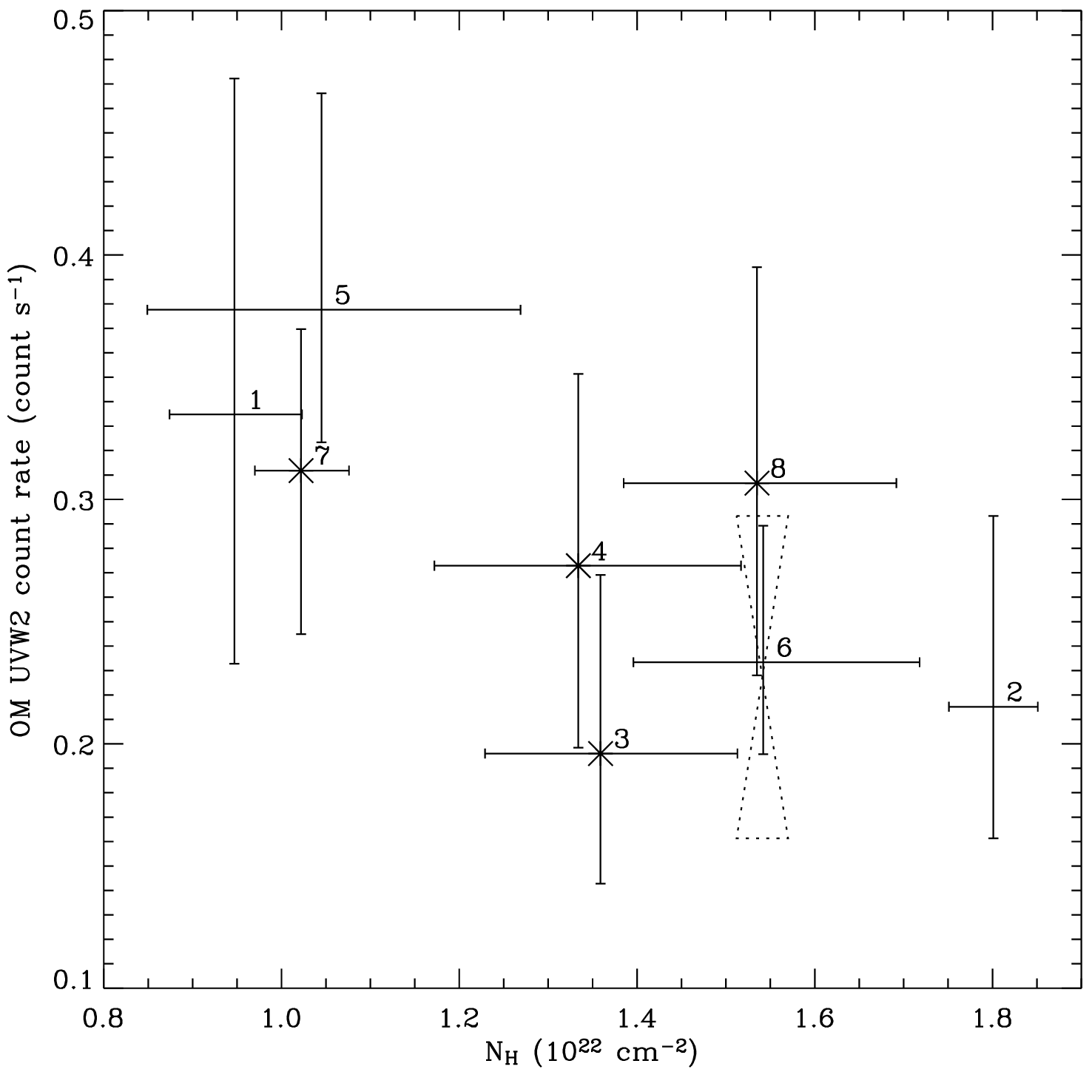}
 \caption{Average OM count rate versus column density. Labels indicate
 \xmm~observations. The crosses mark observations obtained outside the
 optical eclipse. The dotted hourglass shows the column density of
 observation \#2 when using only one-temperature plasma.}
\label{fig:uv_x}
\end{figure*}
}

The online Fig.~\ref{fig:uv_x} shows the scattered plot of the average
OM count rate versus the column density, where the count rate error
include the observed variations of the UV flux (see
Fig.~\ref{fig:om_epic_lightcurves}). The correlation coefficient of
this sample is $-0.72$, suggesting a true correlation between the two
physical parameters as argued by SR, who used smaller error bars for
the count rates. However, we note that the data obtained outside the
eclipse are located within the cloud of points, with error bars
covering all the range of observed values of column density and count
rate. Therefore, the column density variation cannot be attributed to
the disk warp.

Assuming a frequency of one X-ray flare per 650\,ks, as observed in young
solar-mass stars \citep{wolk05}, our 38.8\,h-exposure should detect
only 0.2 flare from \at. The center of the optical eclipse is limited
to about 0.2 in phase. Therefore, the combined probability to observe
by chance an X-ray flare during the center of the transit of an
accretion funnel flow is only 0.04. Moreover, the median flare level in young
solar-mass stars is only 3.5 times the characteristic level
\citep{wolk05}, whereas we observed a flare with a larger
amplitude. This suggests that this event is associated with the
magnetic area corresponding to the base of the dipolar magnetic field
line, which controls the accretion funnel flow.

%________________________________________________________________
%________________________________________________________________
\section{Discussion and conclusions}
\label{discussion}

The high throughput of \xmm~allows us to obtain spectra of \at~at each
phase of its activity. By contrast to previous {\sl ROSAT} observations, 
where the measurement for this source was limited to count rate and
hardness ratios in the soft X-ray energy
band\footnote{To convert the {\sl ROSAT} count rate of \at~to X-ray
  luminosity, \citet{neuhaeuser95} estimate an energy conversion
  factor from the visual extinction (converted to foreground hydrogen
  column density) and the second hardness ratio (sensitive to the
  plasma temperature). This leads to an X-ray luminosity of
  $0.4\times10^{30}$\ergs~\citep[see Table~1 of][]{johnskrull07},
  which is likely underestimated by a factor of ten (see our
  simulations of count rates in Sect.~\ref{archives}), because the column
  density of \at~(see
  Table~\ref{table:fit_epic_wabs_vapec_angr_xest_parameters}, and
  discussion below) is about 10 times larger than the one assumed by
  \citet{neuhaeuser95}.}, the plasma parameters can be derived from
spectral fitting, which provides, in particular, a better measurement
of the column density and the X-ray luminosity.

We find that the column density, derived from the photoelectric
absorption of the X-rays emitted by the active corona of \at, varies
from $N_{\rm H}\sim 1.0 \times 10^{22}$\,cm$^{-2}$ to $1.8 \times
10^{22}$\,cm$^{-2}$. However, the optical extinction of \at~by the
dust is very low with $A_{\rm V}=0.78$\,mag \citep{bouvier99}, which
can be converted to $N_{\rm H} \sim (1.2\pm0.1) \times
10^{21}$\,cm$^{-2}$, using the relation $N_{\rm H}/A_{\rm J} = (5.6
\pm 0.4) \times 10^{21}$\,cm$^{-2}$\,mag$^{-1}$ \citep{vuong03},
combined with the extinction law of \citet{rieke85}, $A_{\rm
  J}=0.282\times A_{\rm V}$ (for their adopted $R_{\rm V}$ value of
3.1). We conclude that there is an excess of column density, varying
from $0.9 \times 10^{22}$ to $1.7 \times 10^{22}$\,cm$^{-2}$. These
values are consistent with the one reported by SR. To explain this
excess of column density, SR introduced an additional absorption in a
disk wind (with no dust, and cool enough to avoid producing any soft
X-ray emission), or a peculiar dust grain distribution with $R_{\rm V}
\sim 0.4$ to reconcile the observed extinction and X-ray absorption.
However, we note that three-dimensional MHD simulations of disk
accretion to a rotating magnetized star \citep[e.g.,][]{romanova03}
show that the stellar magnetosphere is far to be an empty cavity. 
Matter accretes mainly through narrow funnel loaded with high-density
material ($\sim$$10^{12}$\,cm$^{-3}$), which are surrounded by lower
density funnel flows that blanket nearly the whole
magnetosphere. Therefore, we propose to identify this excess of gas
with this lower density funnel flows filling the stellar
magnetosphere. Dividing the excess column density by the width of the
magnetosphere of \at~(about 7.8 stellar radii) leads to a density of a
few $10^{10}$\,cm$^{-3}$, which is compatible with this
interpretation. Moreover, the multiple spirals visible in the simulated
accretion flows should help to produce a variable column density. This low
density gas, located below the radius of dust sublimation (close to
the inner accretion disk for \at), is then dust free.

Accreting TTSs show in average a $\lxlbol$ ratio 2.5 times
smaller than in non-accreting TTSs (see for COUP and XEST results,
\citealt{preibisch05b} and \citealt{briggs07}, respectively). This is
generally interpreted as a direct or indirect consequence of the
accretion process \citep[see discussion
  in][]{preibisch05b,telleschi07,gregory07}. 
The median value of $\log(\lxlbol)$ for \at~is $-3.5$; for comparison
this value is intermediate between $-3.7$ and $-3.3$, the median
values found for accretors and non-accretors, respectively
\citep{preibisch05b,briggs07}.  The peculiar orientation of \at~may
help to overcome partly the extinction by the accretion funnels
\citep{gregory07}.

Recently, \citet{johnskrull07} reported new magnetic field
measurements for CTTSs, based on Zeeman broading of photospheric
absorption lines in the near-IR, showing that the observed mean
magnetic field is in all cases greater than the field predicted by
pressure equipartition arguments. \citet{johnskrull07} suggests
that the very strong fields decrease on these stars the efficiency
with which convective gas motions in the photosphere can tangle
magnetic loops in the corona. For \at, \citet{johnskrull07} reports a
value of 2.78\,kG for the mean magnetic field, that is 2.7 times
larger than the field strength at pressure equipartition; and predicts from
solar scaling an (unobserved) X-ray luminosity of about
$4\times10^{30}$\,\ergs~(for comparison this value is
10 times greater than the minimum level that we observed). 
However, our observation shows that a bright X-ray flare can occur
during the transit of an accretion funnel flow, and locate the active
X-ray area likely close to the foot of the accretion funnel flow. We
speculate that a magnetic interaction exists between the free falling
accretion flow and the strong magnetic field of the stellar corona,
which may trigger magnetic reconnections and give rise to bright X-ray
flares, which boost the X-ray emission. 

This campaign of observations of \at~with \xmm~shows that X-ray
spectroscopy with CCD provides a unique tool to probe the
circumstellar dust-free gas in this object, and that some magnetic
flares are likely associated with the accretion process. Longer
coordinated optical and X-ray/UV observations of \at~are still needed to
obtain continuous light curves, which is crucial to confirm the
flaring behaviour of active regions associated with the accretion
funnel flow, and daily modulation of the UV flux. Simultaneous
Zeeman-Doppler images would help to derive a surface magnetogram; this
map of the magnetic active regions on the stellar photosphere,
combined with the X-ray light curves, would then allow to build a
self-consistent three-dimensional model of the corona of \at. 

Taking into account that the duration of the visibility window for
\at~is about 130\,ksec per \xmm~orbit (2 days), a full monitoring of
two optical eclipse periods of \at~(8.22 days) with \xmm~is equivalent
to an effective exposure of 1.1\,million seconds. This project has the
typical duration of the large programs which are anticipated with
\xmm~for the next decade.

\begin{acknowledgements}
This research is based on observations obtained with \xmm, an ESA
science mission with instruments and contributions directly funded by
ESA Member States and NASA. M.F.\ was supported by the Spanish grants
AYA2004-05395 and AYA2004-21521-E. This research was partly based on
data obtained at the 1.5 m telescope at the Sierra Nevada Observatory,
which is operated by the Consejo Superior de Investigaciones
Cient{\'\i}ficas through the Instituto de Astrof{\'\i}sica de
Andaluc{\'\i}a.
\end{acknowledgements}

\phantomsection 
\addcontentsline{toc}{chapter}{References}
\bibliographystyle{aa}
\bibliography{biblio}

%%%%%%%%%%%%%%%%%%%%%%%%%%%%%%%%%%%%%%%%%%%%%%%%%%%%%%%
% online material
%%%%%%%%%%%%%%%%%%%%%%%%%%%%%%%%%%%%%%%%%%%%%%%%%%%%%%%

\Online
\phantomsection 
\addcontentsline{toc}{chapter}{Online material}
                                                   
%%%%%%%%%%%%%%%%%%%%%%%%%%%%%%%%%%%%%%%%%%%%%%%%%%%%%%%
% Appendix
%%%%%%%%%%%%%%%%%%%%%%%%%%%%%%%%%%%%%%%%%%%%%%%%%%%%%%%

\phantomsection 
\addcontentsline{toc}{chapter}{\appendixname}

\begin{appendix}

%________________________________________________________________
\section{Elemental abundances of the coronal plasma and the absorbing
  material}
\label{appendix:newabun}

The typical values of plasma element abundances observed in the
coronae of young stars with fine X-ray spectroscopy \citep{guedel07},
that we use for the {\tt vapec} coronal plasma model in our X-ray
spectral fitting with {\tt XSPEC}, are given in Col.~(2)--(4) of
Table~\ref{table:xest_angr_aspl_abund}.

The column density of the absorbing material located on the line of
sight is estimated from spectral fitting using a photoelectric
absorption model, that uses photoionization cross sections, and solar
abundances for the material composition. The {\tt wabs} photoelectric
absorption model use photoionization cross sections of \citet{morrison83}, and
(old) solar abundances of \citet{anders89}. Significant revisions of the
solar abundances have been made recently, as a result of the application
of a time-dependent, 3D hydrodynamical model of the solar atmosphere,
instead of 1D hydrostatic models. This has decreased the metal
abundances, in particular of carbon and oxygen, which are the main
contributor to the photoionization cross section above 0.3 and
0.6\,keV, respectively. Consequently, the absolute value of the column
density is dependent of the adopted photoelectric absorption
model. Using updated solar abundances is then crucial when the
absolute value of the column density is needed \citep[e.g.,][]{vuong03}.
Col.~(5)--(7) of Table~\ref{table:xest_angr_aspl_abund}
give the recent compilation of elemental solar abundances by
\cite{asplund05}, where the decrease of metal by comparison with
\citet{anders89} is indicated in the last column.

\begin{figure}[!ht]
\centering
\includegraphics[angle=0,width=0.7\columnwidth]{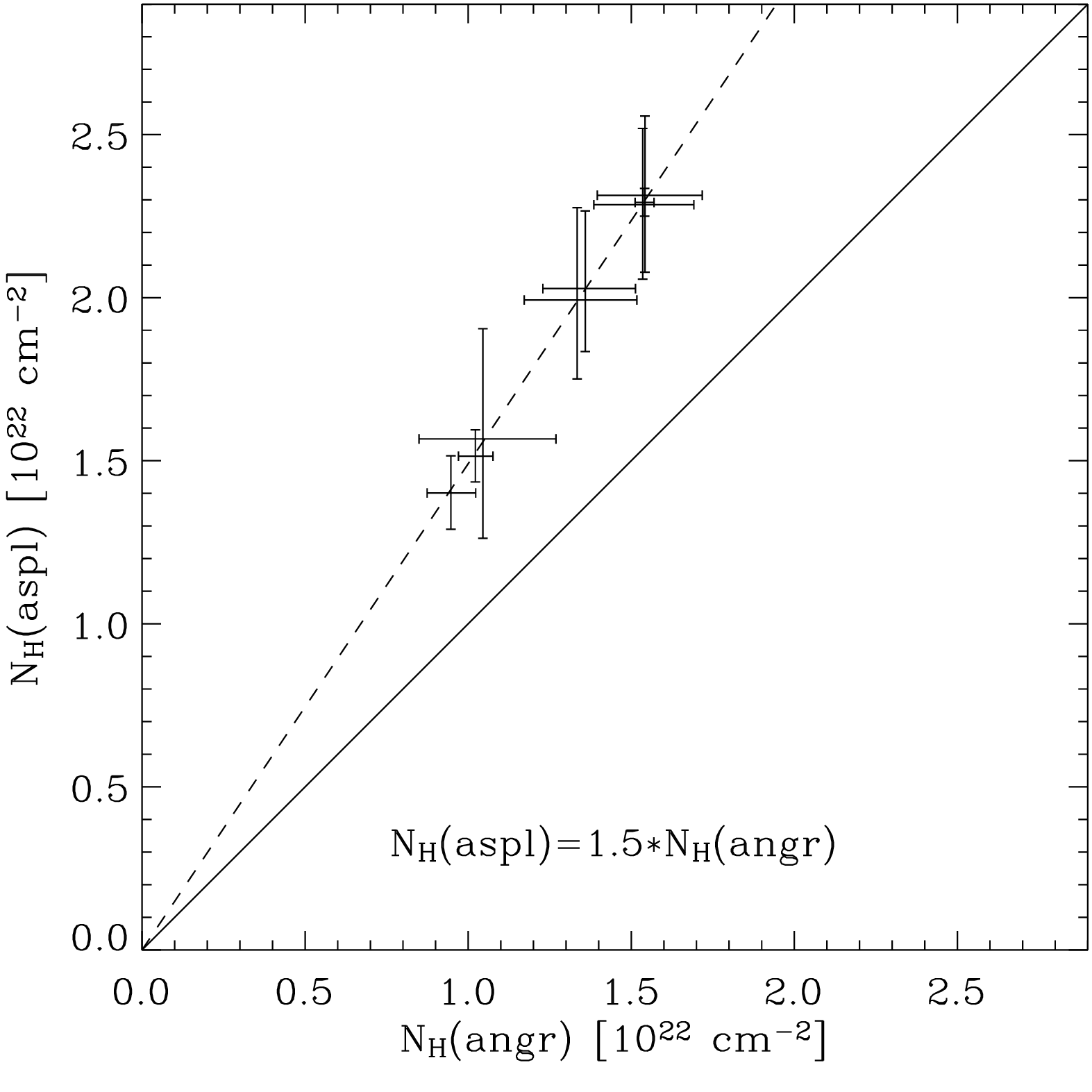}
\caption{Comparison of the column density values obtained from
  spectral fitting when using the old \citep{anders89} and the updated
  \citep{asplund05} solar abundances. The dashed line shows the mean
  average between the two values of column density.}
\label{fig:aspl_vs_angr}
\end{figure}

We test the impact of these udpated solar abundances on our fitting by
replacing {\tt wabs} by {\tt tbvarabs} \citep{wilms00}, which allows
to input new abundances for the absorbing material. Moreover, {\tt
  tbvarabs} uses also updated photoionization cross sections. We find
nearly identical values of temperature and emission measure, however,
as anticipated the column density value is
increased. Fig.~\ref{fig:aspl_vs_angr} shows that column density
values obtained with {\tt wabs} are underestimated by about 50\%.

\begin{table}[!h]
\caption{Elemental abundances of the coronal plasma and the
  absorbing material.}
\label{table:xest_angr_aspl_abund}
\begin{tabular}{@{}crccrcc@{}}
\hline
\hline
\noalign{\smallskip}
   & \multicolumn{3}{c}{Coronal plasma$^\mathrm{a}$} &
\multicolumn{3}{c}{Absorbing material$^\mathrm{b}$} \\
   & \multicolumn{3}{c}{\hrulefill} & \multicolumn{3}{c}{\hrulefill} \\
El & \multicolumn{1}{c}{$A({\rm El})$} & $n({\rm El})/n({\rm H})$ &
angr & \multicolumn{1}{c}{$A({\rm El})$} & $n({\rm El})/n({\rm H})$ & angr \\ 
(1) & (2) & (3) & (4) & (5) & (6) & (7)  \\ 
\hline
\noalign{\smallskip}
He &  10.99 &	9.77E-02 &  1.000 &  10.93 &   8.51E-02 &  0.871 \\
 C &   8.21 &	1.63E-04 &  0.450 &   8.39 &   2.45E-04 &  0.676 \\
 N &   7.95 &	8.83E-05 &  0.788 &   7.78 &   6.03E-05 &  0.538 \\
 O &   8.56 &	3.63E-04 &  0.426 &   8.66 &   4.57E-04 &  0.537 \\
Ne &   8.01 &	1.02E-04 &  0.832 &   7.84 &   6.92E-05 &  0.562 \\
Na &\dotfill&	 \dotfill&\dotfill&   6.17 &   1.48E-06 &  0.691 \\
Mg &   7.00 &	9.99E-06 &  0.263 &   7.53 &   3.39E-05 &  0.892 \\
Al &   6.17 &	1.47E-06 &  0.500 &   6.37 &   2.34E-06 &  0.795 \\
Si &   7.04 &	1.10E-05 &  0.309 &   7.51 &   3.24E-05 &  0.912 \\
 S &   6.83 &	6.76E-06 &  0.417 &   7.14 &   1.38E-05 &  0.852 \\
Cl &\dotfill&	 \dotfill&\dotfill&   5.50 &   3.16E-07 &  1.682 \\
Ar &   6.30 &	2.00E-06 &  0.550 &   6.18 &   1.51E-06 &  0.417 \\
Ca &   5.65 &	4.47E-07 &  0.195 &   6.31 &   2.04E-06 &  0.892 \\
Cr &\dotfill&	 \dotfill&\dotfill&   5.64 &   4.37E-07 &  0.902 \\
Fe &   6.96 &	9.13E-06 &  0.195 &   7.45 &   2.82E-05 &  0.602 \\
Co &\dotfill&	 \dotfill&\dotfill&   4.92 &   8.32E-08 &  0.967 \\
Ni &   5.54 &	3.47E-07 &  0.195 &   6.23 &   1.70E-06 &  0.954 \\
\noalign{\smallskip}
\hline
\end{tabular}
Notes: Col.~(2) and (5) give the element abundances on the logarithmic
astronomical scale, where the numbers of hydrogen atoms are set to
$A({\rm H})=\log n({\rm H})=12$. The numbers of element atoms
normalized to the number of hydrogen atoms are given in Col.~(3) and
(6); Col.~(4) and (7) compare this ratio to \citet{anders89}'s
photospheric abundances.
\begin{list}{}{}
\item[$^{\mathrm{a}}$] Elemental abundances observed in the coronae of young stars with fine X-ray spectroscopy \citep{guedel07} used in {\tt vapec}. 
\item[$^{\mathrm{b}}$] Elemental solar abundances of \citet{asplund05}
  used in {\tt tbvarabs}.
\end{list}
\end{table}
%________________________________________________________________

\end{appendix}

\end{document}